\PassOptionsToPackage{table}{xcolor}
\documentclass[journal]{vgtc}                

\ifpdf
  \pdfoutput=1\relax                   
  \usepackage{graphicx}                
  \DeclareGraphicsExtensions{.pdf,.png,.jpg,.jpeg} 
\else
  \usepackage{graphicx}                
  \DeclareGraphicsExtensions{.eps}     
\fi%

\usepackage{multirow}

\graphicspath{{figures/}{pictures/}{images/}{./}} 

\usepackage{microtype}                 
\PassOptionsToPackage{warn}{textcomp}  
\usepackage{textcomp}                  
\usepackage{mathptmx}                  
\usepackage{cite}                      
\usepackage{tabu}                      
\usepackage{booktabs}                  
\usepackage{cite}
\usepackage{balance}  
\usepackage{graphics} 
\usepackage[T1]{fontenc}
\usepackage{txfonts}
\usepackage{mathptmx}
\usepackage[htt]{hyphenat}
\usepackage{array}
\usepackage{booktabs}
\usepackage{pifont}
\usepackage{xspace}
\usepackage{setspace}
\usepackage{titlesec}
\usepackage{graphicx}
\usepackage[skip=5pt,font=small]{caption}%
\usepackage{microtype} 
\usepackage{ccicons}  
\usepackage{verbatim}
\usepackage{relsize}
\usepackage{etoolbox}
\usepackage{lipsum}   
\usepackage{setspace} 
\usepackage[normalem]{ulem}
\usepackage{enumitem}
\usepackage{relsize,etoolbox}
\usepackage{makecell}

\AtBeginEnvironment{quote}{\small}
\newenvironment{denselist}{
    \begin{list}{\small{$\bullet$}}%
    {\setlength{\itemsep}{0ex} \setlength{\topsep}{0ex}
    \setlength{\parsep}{0pt} \setlength{\itemindent}{0pt}
    \setlength{\leftmargin}{1.5em}
    \setlength{\partopsep}{0pt}}}%
    {\end{list}}
\newcommand{\squishlist}{
   \begin{list}{$\bullet$}
    { \setlength{\itemsep}{0pt}
      \setlength{\parsep}{2pt}
      \setlength{\topsep}{0pt}
      \setlength{\partopsep}{0pt}
      \leftmargin=25pt
\rightmargin=0pt
\labelsep=5pt
\labelwidth=10pt
\itemindent=0pt
\listparindent=0pt
\itemsep=\parsep
    }
}

\newcommand{\squishend}{\end{list}}
\newcommand{\npar}{\par\noindent}
\newcommand{\eat}[1]{}

\newcommand{\techreport}[1]{}

\newcommand{\annon}[1]{}
\newcommand{\boldpara}[1]{\par\noindent\textbf{#1}}
\newcommand{\stitle}[1]{\noindent\textbf{#1}}
\newcommand{\sectitle}[1]{\vspace*{2pt}\noindent\textbf{#1}}
\newcommand{\nstitle}[1]{\\\noindent\textbf{#1}}
\newcommand{\rot}[1]{\rotatebox[origin=c]{90}{\parbox{2.6cm}{\centering#1}}}
\newcommand{\A}{\textbf{A: }}
\newcommand{\M}{\textbf{M: }}
\newcommand{\G}{\textbf{G: }}
\newcommand{\problemlist}{
    \nstitle{Motivating Use Case Challenges:}
    \vspace{-5pt}
    \begin{enumerate}[label=\textbf{C\arabic*:},leftmargin=0.7cm]
}
\newcommand{\featurelist}{
    \stitle{Instantiated Feature:}
    \vspace{-5pt}
    \begin{enumerate}[label=\textbf{F\arabic*:},leftmargin=0.7cm]
}
\newcommand{\enumend}{
  \end{enumerate}
  \vspace{-5pt}
}

\def\pprw{8.5in}
\def\pprh{11in}

\usepackage{xcolor}
\newcommand{\zv}{Zenvisage\xspace}
\newcommand{\zvpp}{\textit{zenvisage++}\xspace}

\newcommand{\ccut}[1]{} 
\newcommand{\cchange}[1]{#1}
\newcommand{\cut}[1]{}
\newcommand{\rchange}[1]{#1}
\newcommand{\bartext}[1]{{\small \textsf{#1}}}

\usepackage{xparse}   

\onlineid{1041}
\ieeedoi{10.1109/TVCG.2019.2934666}
\vgtccategory{Research}
\vgtcpapertype{application/design study}

\title{\emph{You can't always sketch what you want}: \\ Understanding Sensemaking in Visual Query Systems}

\author{Doris Jung-Lin Lee, John Lee, Tarique Siddiqui, Jaewoo Kim, Karrie Karahalios, Aditya Parameswaran}
\shortauthortitle{Lee \MakeLowercase{\textit{et al.}}: \emph{You can't always sketch what you want}: \\ Understanding Sensemaking in Visual Query Systems}

\abstract{
  Visual query systems (VQSs) empower users to interactively search for line charts with desired visual patterns, typically specified using intuitive sketch-based interfaces. Despite decades of past work on VQSs, these efforts have not translated to adoption in practice, possibly because VQSs are largely evaluated in unrealistic lab-based settings. To remedy this gap in adoption, we collaborated with experts from three diverse domains---astronomy, genetics, and material science---via a year-long \rchange{user-centered} design process to develop a VQS that supports their workflow and analytical needs, and evaluate how VQSs can be used in practice. Our study results reveal that ad-hoc sketch-only querying is not as commonly used as prior work suggests, since analysts are often unable to precisely express their patterns of interest. In addition, we characterize three essential sensemaking processes supported by our enhanced VQS. We discover that participants employ all three processes, but in different proportions, depending on the analytical needs in each domain. \cchange{Our findings suggest that all three sensemaking processes must be integrated in order to make future VQSs useful for a wide range of analytical inquiries.}
}

\renewcommand{\authorfootertext}{\vspace*{-5pt}\item\textit{Doris and Aditya are with University of California, Berkeley. \\Email: \textrm{\{}dorislee, adityagp\textrm{\}}@berkeley.edu\\ \item John, Tarique, Jaewoo and Karrie are with University of Illinois, Urbana-Champaign. \\E-mail: \textrm{\{} lee98, tsiddiq2, jkim475, kkarahal\textrm{\}}@illinois.edu}}

\vgtcinsertpkg

\begin{document}
\maketitle
\raggedbottom
 \section{Introduction\label{sec:intro}}
 Line charts are commonly employed during data exploration---the intuitive connected patterns often illustrate complex underlying processes and yield interpretable and visually compelling data-driven narratives~\cite{Few2012}. However, discovering line charts that display certain meaningful patterns, trends, or characteristics of interest is often  an overwhelming and error-prone process, consisting of manual examination of large numbers of line charts. For example, when trying to find supernovae, which exhibits a unique pattern of brightness over time (an initial peak followed by a long-tail decay), astronomers often have to manually construct and inspect thousands of line chart visualizations to find ones with their desired pattern. To address this exploration challenge, there has been a large number of papers dedicated to building \emph{Visual Query Systems} (VQSs)---a term coined by Ryall et al.~\cite{ryall2005querylines} to describe systems that allow users to specify and search for desired line chart patterns via visual interfaces~\cite{mohebbi2011google,Hochheiser2004,wattenberg2001sketching,Siddiqui2017VLDB,ryall2005querylines,correll2016semantics,Mannino2018,Eichmann2015,Holz2009}. These interfaces typically include a sketching canvas where users can draw a pattern of interest, with the system automatically traversing all potential visualization candidates to find those that match the specification.  
 \par While these intuitive specification interfaces were proposed as a promising solution to the problem of painful manual exploration of visualizations for time-series analysis~\cite{ryall2005querylines,wattenberg2001sketching}, to the best of our knowledge, VQSs have not lived up to these expectations and are not very commonly used in practice. One likely reason for the lack of VQS adoption may be attributed to how prior work \rchange{has} focused almost exclusively on optimizing the pattern-matching algorithms and interactions, with few invested in understanding actual user needs and how VQSs can be used for solving real-world problems. {\em Our paper seeks to understand how VQSs can actually be used in practice, as a first step towards the broad adoption of VQSs in data analysis}. Unlike prior work on VQSs, we set out to not only evaluate VQSs in-situ on real problem domains, but also involve participants from these domains in the VQS design. We present findings from a series of interviews, contextual \rchange{inquiry}, participatory design, and user studies with scientists from three different domains---{\em astronomy, genetics,} and {\em material science}---over the course of a year-long collaboration. The amount of time we invested in each of these three diverse domains surpasses the norm in this field and is key to uncovering the insights presented in this paper. \cut{As illustrated in Figure~\ref{science_goal},}These domains were selected to capture a diverse set of goals and datasets wherein VQSs can help address important scientific questions, such as: How does a treatment affect the expression of a gene in a breast cancer cell-line? Which battery components have sustainable levels of energy-efficiency and are safe and cheap to manufacture in production?
 \begin{figure}
    \centering
    \includegraphics[width=0.85\linewidth]{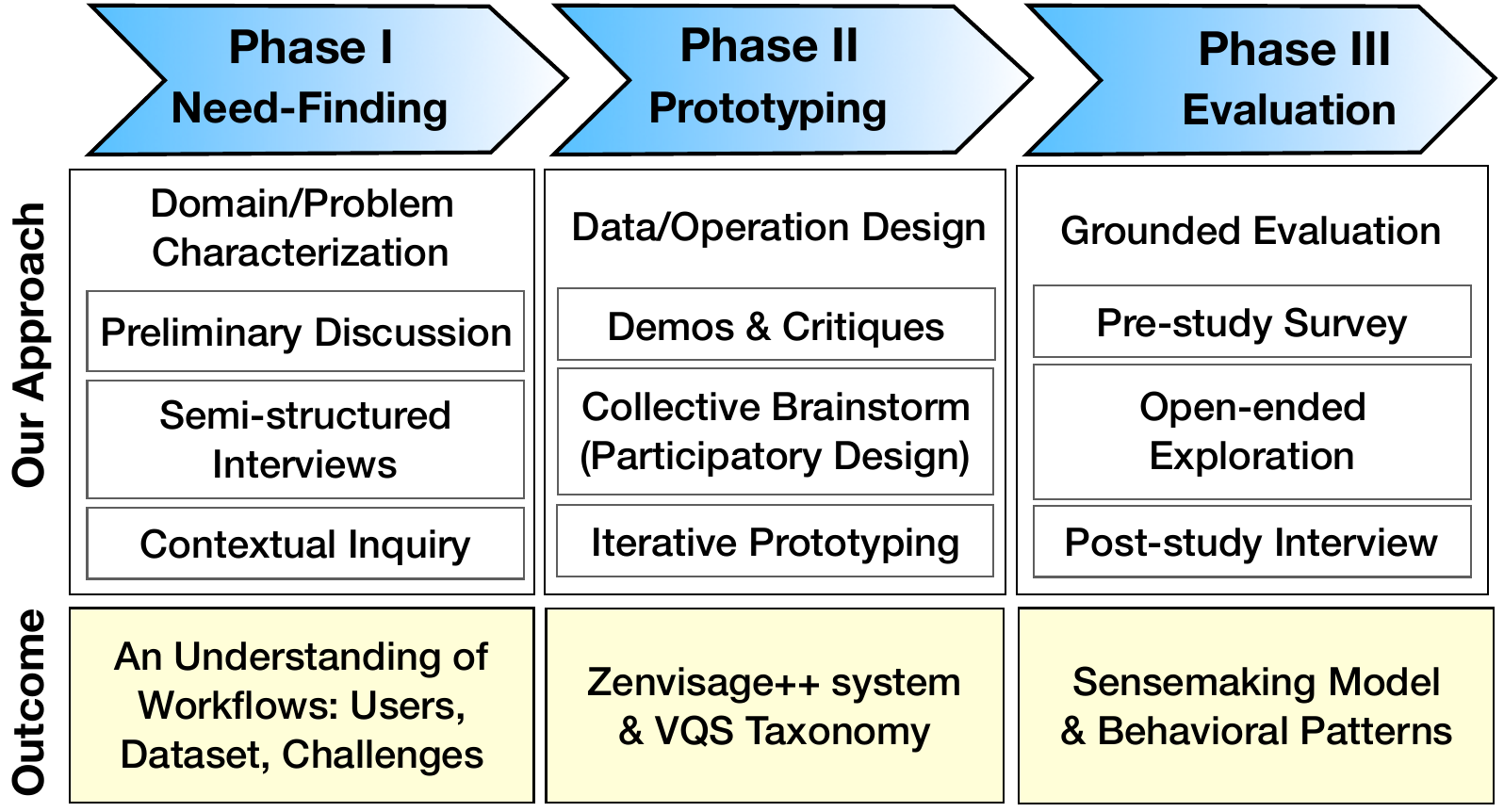}
    \caption{Lifecycle model summarizing our research approach and the outcome of each phase.}
    \label{methodFlowchart}
    \vspace*{-20pt}
  \end{figure}
 \par \rchange{In this work, we adapt methods from user-centered design (UCD)~\cite{Norman1986,Nielsen1994,Gould1983}, such as interviews, contextual inquiry, and participatory design, into our design-implementation-evaluation cycle~\cite{Sharp2007}; our methodology is summarized in Figure~\ref{methodFlowchart}.} Via contextual \rchange{inquiry} and interviews, we first identified challenges in existing data analysis workflows in these domains that could be potentially addressed by a VQS. Building on top of an existing open-source VQS, \zv~\cite{Siddiqui2017,Siddiqui2017VLDB}, we \rchange{iterated on the design of the VQS with participants over the course of a year to} better compose data exploration workflows that lead to insight discovery. Rather than targeting a domain-specific solution, we \rchange{engaged with} multiple domains \cut{(an uncommon practice in visualization design studies) }to observe differences and commonalities across domains and synthesize high-level insights regarding the use of VQSs. While \rchange{conducting} this multi-phased, mixed-methods research agenda across three diverse use cases was challenging, this endeavor was necessary for addressing the qualitative, participant-centered research questions investigated.
 \par We organize our \cut{PD}\rchange{design study} findings into a taxonomy of VQS capabilities, involving three sensemaking processes inspired by Pirolli and Card's notional model of analyst sensemaking~\cite{Pirolli}. The sensemaking processes include \emph{top-down pattern search} (translating a pattern ``in-the-head'' into a visual query), \emph{bottom-up data-driven inquiries} (querying or recommending based on data), and \emph{context-creation} (navigating across different collections of visualizations). We find that prior VQSs have focused on enabling top-down processes (via sketching capabilities), but have largely overlooked the two other processes that we found to be essential in all three domains. These missing capabilities partially \rchange{explain} why prior VQSs have not been widely adopted in practice.
 \cut{
   \begin{figure}[ht!]
    \centering
    \includegraphics[width=0.9\linewidth]{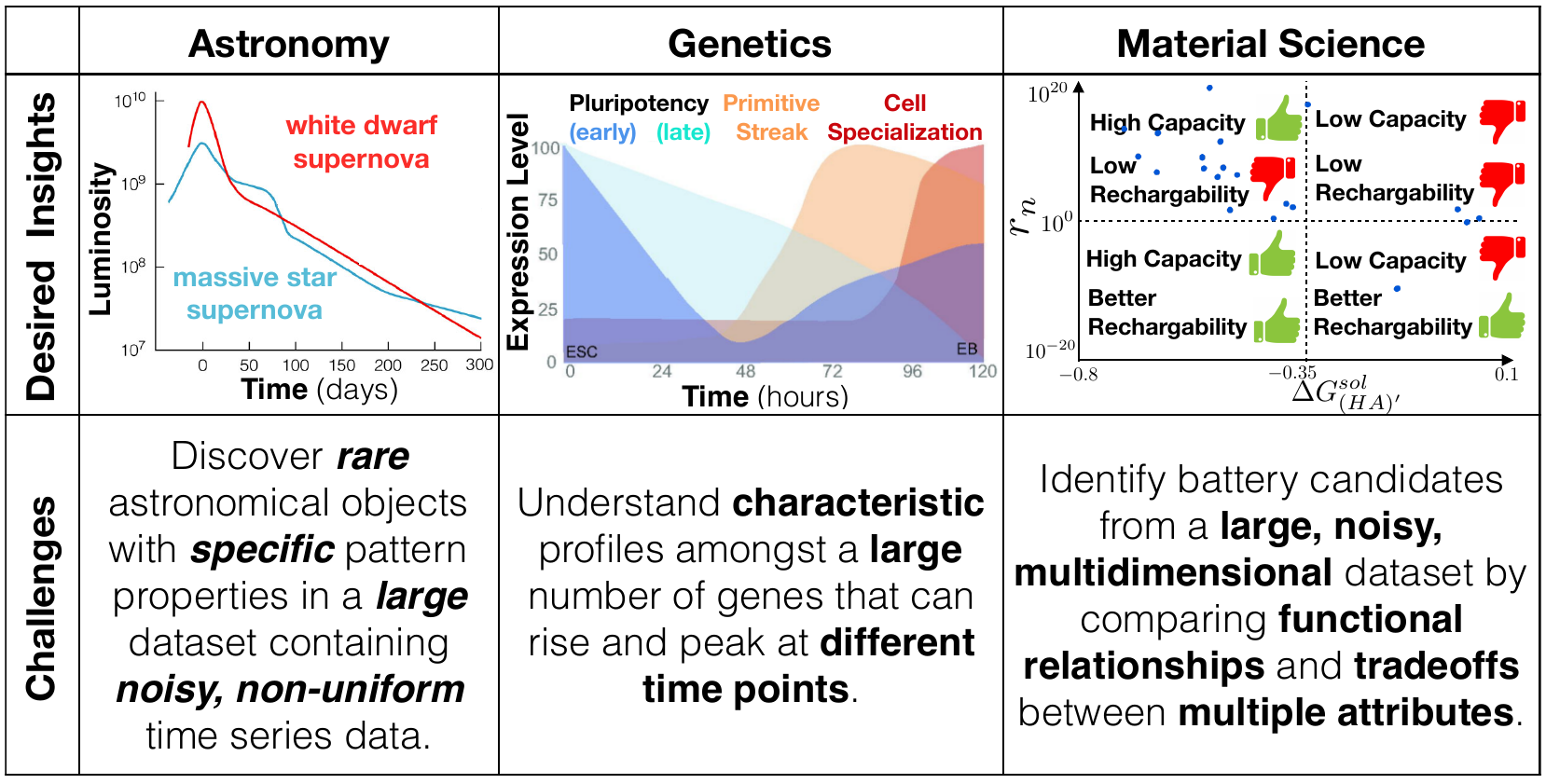}
    \caption{Desired insights, problem and dataset challenges for each of the three application domains in our study.}
    \label{science_goal}
    \vspace*{-20pt}
   \end{figure}
 }
 \par We \rchange{finally conducted an} evaluation study with nine participants using our final VQS prototype to address their research questions on their own datasets. During this study, participants gained novel scientific insights,
 such as identifying a star that was known to harbor a Jupiter-sized planet, discovering a previously-unknown relationship between solvent properties, and finding characteristic gene expression profiles confirming the results of a related publication. 
 \par \rchange{During this evaluation study}, we were somewhat surprised to discover that sketching a pattern for querying is often ineffective on its own. This is due to the fact that sketching makes the assumption that users know the pattern that they want to sketch and are able to sketch it precisely. However, this is \rchange{typically} not the case in practice. For example, the geneticists from our study often did not have a preconceived knowledge of what to sketch for and relied heavily on VQS-recommended common and outlying patterns to jumpstart their queries. Likewise, while the material scientists from our study were interested in datapoints that fall within specific value-ranges, they did not have an apriori notion of what their desired patterns would look like. Overall, participants typically opted to combine sketching with other means of pattern specification---one common mechanism was to drag-and-drop a recommended pattern onto the canvas, and then modify it (e.g., by smoothing it out). 
 \par To further understand how participants engaged with VQSs in their analytical workflows, we constructed a Markov model to characterize how participants transitioned between different sensemaking processes during their analysis. We found that participants often constructed a diverse set of analytical workflows tailored to their domains by focusing around a primary sensemaking process, while iteratively interleaving their analysis with the two other processes. This finding points to how all three sensemaking processes, along with seamless transitions between them, are crucial for enabling the effective use and adoption of VQSs for addressing real-world challenges.
 \par To the best of our knowledge, our study is the \emph{first to holistically examine how VQSs can be designed to fit the needs of real-world analysts, and how they are actually used in practice}. Working with participants from multiple domains enabled us to compare the differences and commonalities across different domains, thereby identifying general VQS challenges and requirements for supporting common analytical goals\cut{, as illustrated in Figure~\ref{science_goal}}. Our contributions include:
 \begin{denselist}
 \item a characterization of the problems addressable by VQSs through design studies with three different domains,
 \item a taxonomy of essential VQSs capabilities, leading to a sensemaking model for VQSs\cut{, grounded in participatory design findings}, 
 \item an integrative VQS, \zvpp\cut{, post participatory design,} capable of facilitating rapid hypothesis generation and insight discovery, \rchange{resulting from iteration with end-users}, 
 \item study findings on how VQSs are used in practice, leading to the development of a novel sensemaking model for VQSs. 
 \end{denselist}
 Our work not only opens up a new space of opportunities beyond the narrow use cases considered by prior studies, but also advocates common design guidelines and end-user considerations for building next-generation VQSs.
  \section{Related Work\label{sec:relatedworks}}
  \npar We will now describe past work in visual query systems and existing evaluation methods of visualization systems to provide background and motivation for our work. \cut{Then, we will introduce Pirolli and Card's sensemaking model, which serves as a framework for contextualizing our study findings.}
  \par \sectitle{Visual Query Systems: Definition and Brief Survey}
  \npar The term \emph{visual query system} (VQS) was introduced by Ryall et al.~\cite{ryall2005querylines} and Correll and Gleicher~\cite{correll2016semantics} to describe systems that enable analysts to directly search for \rchange{line chart}\cut{time-series\footnote{\cut{In this paper, we use this term to encompass line charts in general, since one of our domains (material science) does not visualize time on the x-axis.}}} visualizations matching a queried pattern, constructed through a visual specification interface. Examples of such systems include TimeSearcher~\cite{Hochheiser2001,Hochheiser2004}, where the query specification mechanism is a rectangular box, with the tool filtering out all of the \rchange{line charts} that do not pass through it, and QuerySketch~\cite{wattenberg2001sketching} and Google Correlate~\cite{mohebbi2011google}, where the query is sketched as a pattern on canvas, with the tool filtering out all of the \rchange{line charts} that have a different shape. Subsequent work, including TimeSketch~\cite{Eichmann2015}, SketchQuery~\cite{correll2016semantics}, and Qetch~\cite{Mannino2018}, recognized the ambiguity in sketching by studying how humans rank similarity in patterns. Finer-grained specification interfaces and pattern-matching algorithms have also been developed to improve the expressiveness of sketched queries and clarify how a sketch should be interpreted. These VQSs include QueryLines~\cite{ryall2005querylines} where queries can be flexibly composed of soft constraints and preferences and SoftSelect~\cite{Holz2009} where users can vary the level of sketch similarity across a search pattern. Beyond sketching, \zv~\cite{Siddiqui2017,Siddiqui2017VLDB}, SketchQuery, and TimeSearcher allow users to submit an existing visualization as the query, either via drag-and-drop or double-clicking on the existing visualization. In our work, we built on our system, \zv, since it was open-source, extensible, and included features beyond \rchange{the pattern match} specification typically found in \rchange{other} systems, \rchange{such as the ability to add data filters and examine recommended patterns~\cite{Siddiqui2017}}.\cut{as compared in Table~\ref{table:relatedwork}.} 
  \par \sectitle{Design and Evaluation Methodologies for Visualization Systems}
  \npar Visualization systems are typically evaluated via in-lab usability studies or controlled studies against existing visualization baselines~\cite{Plaisant2004,North2006,Yi2008}. However, successful lab-tested systems do not always translate to community acceptance and adoption. \cut{For instance, while decades of work have shown VQSs to be effective in controlled lab studies, they have not gained widespread adoption. }\ccut{Unlike our work, past VQSs have never been designed and evaluated in-situ on multiple real-world use cases. Even when use cases were involved~\cite{Hochheiser2004,correll2016semantics}, the inclusion of these case studies served as a post-hoc demonstrative case study that had little influence on the major design decisions of the system.} The unrealistic nature of controlled studies has prompted the visualization research community to develop more participant-centered, ethnographic approaches for understanding how analysts perform visual data analysis and reasoning~\cite{Plaisant2004,lam2012empirical,shneiderman2006strategies,munzner2009nested,Sedlmair2012}. For example, multi-dimensional, in-depth, long-term case studies (MILCs) combine interviews, surveys, logging, and other empirical artifacts to create a holistic understanding of how a visualization system can be used in its intended environment \cite{shneiderman2006strategies}. 
  \par In the VQS literature, even though the development and evaluation of advanced VQS algorithms and interactions has been well underway for many years, prior work has yet to characterize and understand the needs of target users and observe how VQSs may be used as part of a real-world workflow, in order to address the initial questions of: 1) whether the problems that VQSs aim to address are even the right ones to address and 2) whether the chosen operations actually solve the user's problems. In the context of Munzner's nested model for visualization design and evaluation~\cite{munzner2009nested}, this gap between research and adoption stems from the common ``\textit{downstream threat}'' of jumping prematurely into the deep levels of \textit{encoding, interaction, or algorithm design}, before a proper \textit{domain problem characterization} and \textit{data/operation abstraction design} is performed. Our work fills this crucial gap in the existing literature and highlights how incorrect assumptions adopted by most prior work in this space regarding the first two stages of Munzner's model may have led to the present-day failures in VQS adoption.
  \par We performed design studies~\cite{lam2012empirical,shneiderman2006strategies,Sedlmair2012} with three different subject areas for \textit{domain problem characterization} by adopting \rchange{user-centered design practices}. \rchange{User-centered design (UCD)~\cite{Norman1986,Nielsen1994,Gould1983} is a class of techniques for iteratively designing a product that fits the needs and desires of users. In UCD, users convey their needs to inform design decisions. Through participatory design (PD)~\cite{Schuler1993,Muller1993}, we engaged potential stakeholders as active co-designers early on and during every step of the design process, in order to develop a system that they may eventually adopt in their analytical workflows. Participatory design is a well-established UCD approach} in the CHI and CSCW \rchange{communities} and has been successfully applied to develop systems for visual analytics~\cite{Aragon2008,Chuang2012}, tangible museum experiences~\cite{Ciolfi2016}, and scientific collaborations~\cite{Poon2008,Chen2016}. \cut{We chose to perform participatory design over other techniques for usability evaluation (such as long-term field study deployments or formative testing), since our goal was to engage potential stakeholders early on and in every step of the design process to ensure that design decisions are based on actual user needs and that we ultimately develop a system that may eventually be adopted in their analytical workflows.} 
  \par In order to ``\textit{[develop] a system model that will support [the] user's work}'' that subsequently ``\textit{fosters participatory design}'', Holzblatt and Jones~\cite{HoltzblattJones} describe contextual inquiry as a technique where researchers observe \rchange{participants} in their own work environment. Likewise, we first perform \rchange{contextual inquiry} and interviews with participants to understand their research questions and the challenges associated with their existing analytical workflows, and to identify design opportunities for VQSs. \cut{Past research has found that the use of functional prototypes is a common and effective way of engaging with participants, by providing a starting point for participatory design~\cite{Ciolfi2016}. Similarly, we provide a functional prototype at the beginning of the participatory design sessions to showcase the capabilities of VQSs.}\ccut{Since our participants were not aware of the existence of VQSs, let alone using them in their workflows, they would not have been able to imagine use cases for VQS without a starting point.}\cut{Likewise, the use of ``\textit{simulated future work situation}'' (where users are introduced to the envisioned use of the prototype) is prevalent in cooperative prototyping when the real use of the prototype is not feasible~\cite{Grnbak1991}.}To better understand how VQSs can be used in-situ \rchange{in} participant's existing \rchange{workflows}, we regularly gathered feedback from participants and collaboratively envisioned potential designs \rchange{by demonstrating} preliminary versions of our protoype \zvpp. \rchange{Based on our design findings, we contribute to the \textit{data/operation abstraction design} of VQSs in Munzner's model by developing a taxonomy for characterizing how analysts make use of VQSs to accomplish their analytical tasks.} Finally, we validated our abstraction design with grounded evaluation~\cite{Plaisant2004,Isenberg2008}, where participants were invited to bring in their own datasets and research problems that they have a vested interest in to test our final deployed system.
 \cut{
    \par \sectitle{Sensemaking Models for Visual Analytics}
    \npar Based on our participatory design findings, we contribute to the \textit{data/operation abstraction design} of VQSs in Munzner's model by developing a taxonomy for understanding how analysts make use of VQSs to accomplish their analytical tasks. To develop a sensemaking model for VQSs, we draw from Pirolli and Card's \rchange{influential} paper on information sensemaking based on cognitive task analysis of intelligence analysts~\cite{Pirolli}. The sensemaking framework was designed to capture how expert analysts iteratively search and represent gathered evidence into a conceptual model (\emph{schema}). Many papers have applied this sensemaking framework to motivate tool designs, such as for exploratory browsing of visualizations in large datasets~\cite{Battle2016} and of the Web~\cite{Olston2003}. The sensemaking framework has also been used for understanding and modeling user behavior in visual analytics, such as how analysts gain insights from visualizations~\cite{Yi2008}, how biases can be introduced during visual analysis~\cite{Wall2017}, and how analysts transition between natural-language data facts and visualizations~\cite{Srinivasan2019}.
    \par In this framework, the sensemaking process can be organized into: 1) a foraging loop that searches for information to further schema organization and 2) a sensemaking loop for constructing a schema that best aligns with the insights obtained from the analysis. Overall, the model distinguishes between information processing tasks that are \textit{top-down} (from theory to data) and \textit{bottom-up} (from data to theory), described more in Section~\ref{sec:sensemaking}. We were inspired by this model for expert intelligence analysis as it bears semblance to our work for studying how domain experts perform exploratory visual analysis using VQSs.
  }
  \section{Methods\label{sec:methods}}
  Via interviews and contextual \rchange{inquiry} in participants' normal work environments, we first identified the needs and challenges in participants' existing data analysis workflows. Given these challenges, we collaboratively designed VQS functionalities by engaging with experts from three different domains \cut{in the process of participatory design}\rchange{throughout the design process}, leading to a final prototype \zvpp. \rchange{After the design phase}, we conducted an evaluation study to understand how VQSs are used in the real-world analytical workflows. \rchange{Our research methodology is illustrated in Figure~\ref{methodFlowchart}; we }now describe the study procedure in more detail.
  \subsection{\rchange{Phase I: Need-finding}}
  \par We recruited participants by reaching out to research groups who have experienced challenges in data exploration, via email and word-of-mouth. Based on early conversations with analysts from 12 different potential application areas, we narrowed down to three use cases in astronomy, genetics, and material science \rchange{through a process similar to the ``\textit{winnow}'' stage in Sedlmair et al.~\cite{Sedlmair2012}. The domains were} chosen based on their suitability for VQSs as well as diversity in use cases. Six scientists\cut{(1 female, 5 male), with an average of more than 6 years of}\rchange{, with extensive} research experience in their respective fields, participated in the design process. \cut{Via interviews and contextual inquiries,}\rchange{We interviewed participants to learn about their dataset and research questions, shadowed participants in conducting their existing analysis workflows, and subsequently discussed the needs and challenges of their use cases. The interviews were semi-structured and focused on how the analytical tasks in their workflows relate to the scientific questions they were interested in.} \cut{We discuss these findings in Section~\ref{sec:participantdatasets}.}
  \subsection{\rchange{Phase II: Collaborative Prototyping}}
  \par For \rchange{iterative prototyping}, we built on top of an existing open-source VQS, \zv~\cite{Siddiqui2017,Siddiqui2017VLDB}, to create a functional prototype \rchange{to showcase the capabilities of \rchange{VQSs}. The use of functional prototypes is a common and effective way of engaging with participants, by providing a starting point for \cut{participatory}\rchange{collaborative design}~\cite{Ciolfi2016}}. \rchange{We} collaborated with each team closely with \cut{an average of}\rchange{approximately} two 1-hour-long meetings per month, where we learned \rchange{more} about their datasets, objectives, and what additional VQS functionalities could help address their research questions. \cut{Since some of the essential features that were crucial for effective exploration were lacking in \zv and still under development in the new version of our VQS, \zvpp, we did not provide a deployed prototype for participants to actively use on their own during the design phase.}\rchange{During these meetings, we collectively brainstormed with participants on the design of the prototype. Participants} also had the opportunity to interact with the prototype through the help of a guided facilitator. \rchange{Through these excercises,} we elicited feedback from participants on how the VQS could better support their scientific goals and identified and incorporated several crucial capabilities into \zvpp.\cut{, described more in Section~\ref{sec:pd_findings}.}
  \cut{A summary timeline of our collaboration with participants over a year can be found in Figure \ref{timeline} in Appendix \ref{apdx:pdartifact}.} 
  \cut{
    \begin{table}[ht!]
    \centering
    \includegraphics[width=0.9\linewidth]{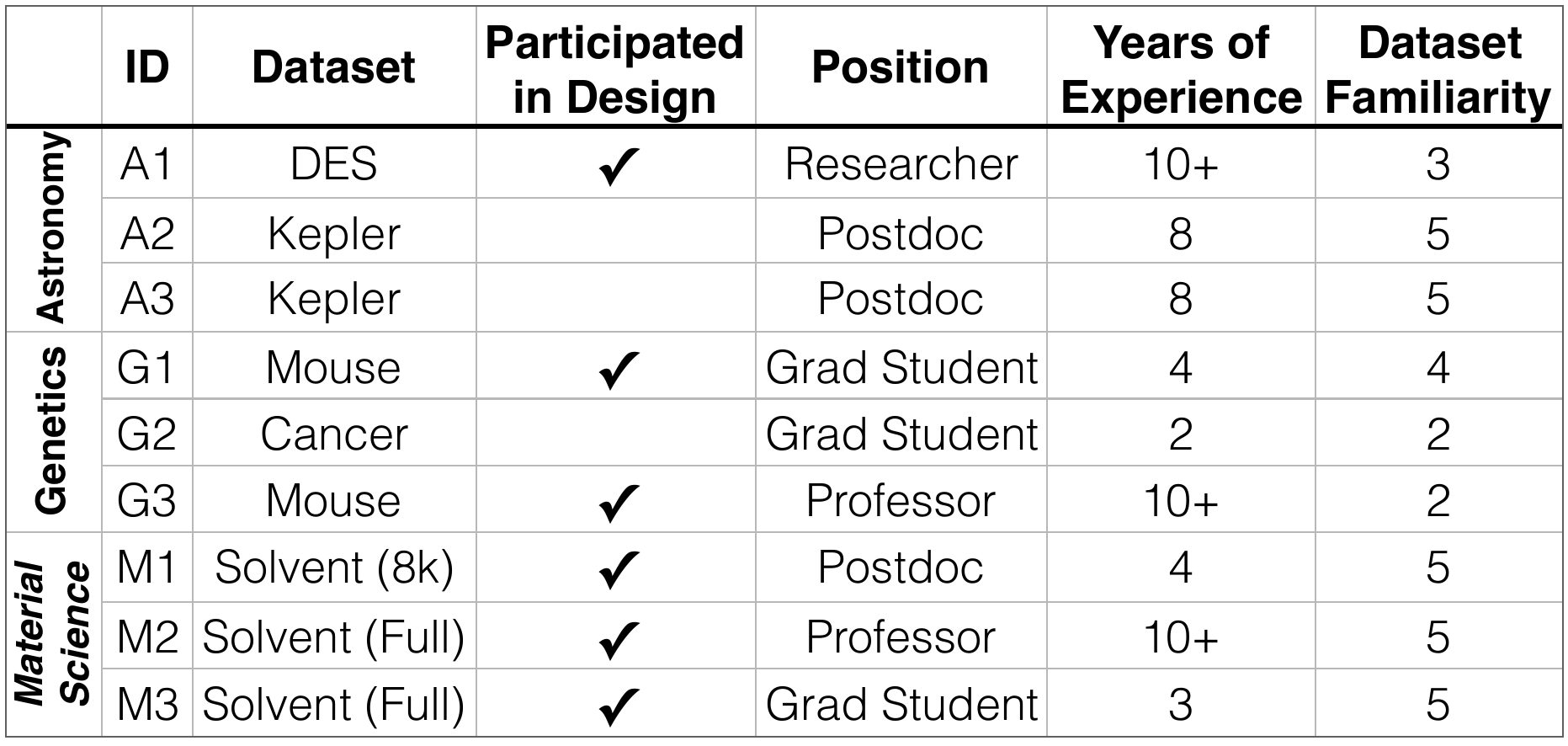}
    \caption{Participant information. The Likert scale used for dataset familiarity ranges from 1 (not familiar) to 5 (extremely familiar).}
    \label{participants}
    \vspace*{-15pt}
  \end{table}
  }
  \subsection{\rchange{Phase III: Grounded Evaluation}}
  \cut{At the end of our \rchange{design study}}\rchange{After the prototyping phase}, we performed a qualitative evaluation to study how analysts interact with different VQS components in practice. Participants used datasets that they have a vested interest in exploring to address unanswered research questions (a total of six different datasets across nine participants). The evaluation study participants included the six scientists from \rchange{Phase I and II}, along with three additional ``blank-slate'' participants who had never encountered \zvpp before\rchange{\footnote{Details regarding participants can be found in the appendix in Table~\ref{participants}.}} The use of all or a subset of the project stakeholders as evaluation participants is typical in participatory design~\cite{Bossen2016}. While the small sample size of participants \rchange{may be viewed as a limitation, this is a pervading challenge when recruiting domain-experts\cite{Batch2018,Mclachlan2008}}\cut{, whose specific expertise and skills are rare and have limited time due to their workplace demands relative to the general population}. Nevertheless, even studies with a small group of domain experts involved are invaluable for understanding expert needs~\cite{Sedlmair2012}. 
  \par Evaluation study participants were recruited from each of the three aforementioned research groups, as well as domain-specific mailing lists. Prior to the study, we asked potential participants to fill out a pre-study survey to determine eligibility. Eligibility criteria included: being an active researcher in the subject area with more than one year of experience, and having worked on a research project involving data of the same nature used in \rchange{the design phase}. \cut{None of the participants received monetary compensation for the study, as this is not a common practice for \rchange{collaborative} design with stakeholders~\cite{Ommen2016,McNally2017}.} \techreport{The research questions and objectives of the participants were diverse even among the same subject area. Examples included understanding gene expression profiles of breast cancer cells after a particular treatment and comparing common patterns among stars that exhibit planetary transits versus stars that do not.\techreport{from the Kepler space telescope\footnote{\url{www.nasa.gov/mission_pages/kepler/main/index.html}}.}}
  \par At the start of the \rchange{in-lab evaluation} study, participants were provided with an interactive walk-through of \zvpp and given approximately ten minutes for a guided exploration of a preloaded real-estate example dataset. \techreport{from Zillow \cite{zillow}. This dataset contained housing data for various cities, metropolitan areas, and states in the U.S. from 2004-15.}After familiarizing themselves with the tool, we loaded the participant's dataset and encouraged them to talk-aloud during data exploration, and use external\cut{ tools or other} resources as needed. If the participant was out of ideas\ccut{ for three minutes}, we suggested one of the main VQS functionalities\cut{\footnote{query by sketching, drag-and-drop, pattern loading, input equations, representative and outliers, narrow/ignore x-range options, filtering, data smoothing, creating dynamic classes,  data export}} that they had not yet used. If \rchange{this operation was} not applicable to their specific dataset, they were allowed to skip the operation after having considered it. The user study \rchange{lasted for about an hour and} ended after they covered all the main functionalities. After the study, we asked participants open-ended questions about their experience.
 \section{Current Participant Workflows and Opportunities\label{sec:participantdatasets}}
 In this section, we describe our study participants, their scientific goals, and their preferred analysis workflows, based on \rchange{Phase I} of our \cut{design }study.\cut{, where we conducted contextual \rchange{inquiry} to learn about participant's existing workflows.} \cut{We will present findings from our collaborative design process in Section~\ref{sec:pd_findings}.}
 While we collaborated with each application domain in depth, we focus on the key findings from each domain to highlight their commonalities and differences, in order to provide a backdrop for our VQS findings described later on. Comparing and contrasting between the diverse set of questions, datasets, and challenges across these three use cases revealed new cross-disciplinary insights essential to better understand how VQSs can be extended for novel and unforeseen use cases.
 \begin{figure*}[ht!]
   \centering
   \vspace{-5pt}
   \includegraphics[width=0.85\linewidth]{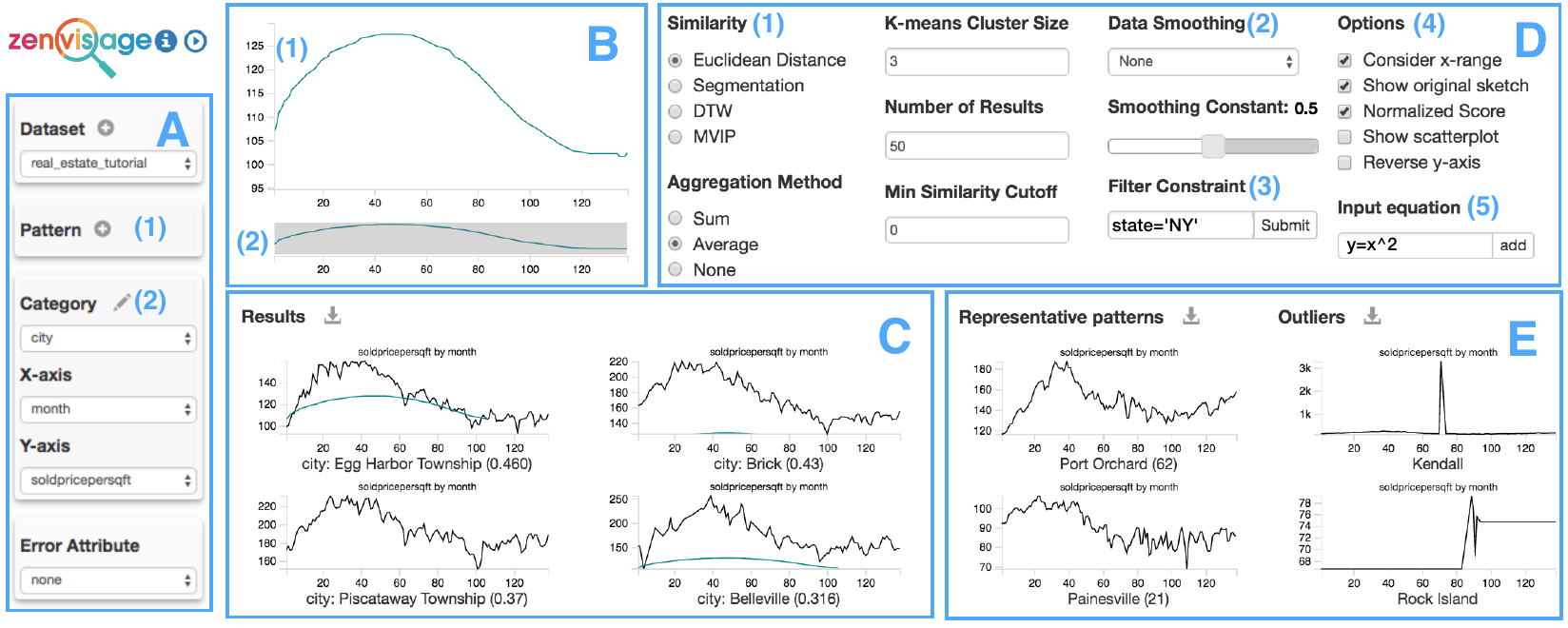} 
   \vspace{-5pt}\caption{The \zvpp system consists of : (A) data selection panel (where users can select visualized dataset and attributes), (B) query canvas (where the queried data pattern is submitted and displayed), (C) results panel (where the visualizations most similar to the queried pattern are displayed as a ranked list), (D) control panel (where users can adjust various system-level settings), and (E) recommendations (where the typical and outlying trends in the dataset is displayed).}
   \label{zvOverview}
   \vspace*{-20pt}
 \end{figure*}
 \subsection{Astronomy}
 \par\sectitle{Participants and Goals:} 
 \npar The Dark Energy Survey (DES) is a multi-institution project that surveys 300 million galaxies over 525 nights to study dark energy~\cite{DrlicaWagner2018}. The telescope used to survey these galaxies also focuses on smaller patches of the sky on a weekly interval to discover astronomical transients, i.e., objects whose brightness changes dramatically as a function of time, such as supernovae or quasars. Their dataset consisted of a large collection of \emph{light curves}: brightness observations over time, one associated with each astronomical object, plotted as a time series. Over five months, we worked closely with A1, an astronomer on the project's data management team at a supercomputing facility. Their scientific goal was to \rchange{\emph{identify potential astronomical transients in order to study their properties}, i.e., identify patterns in line charts}. \cut{These insights can help further constrain physical models regarding the formation of these objects.}
 \cut{
	 \begin{figure}[ht!]
	   \centering
	   \includegraphics[width=\linewidth]{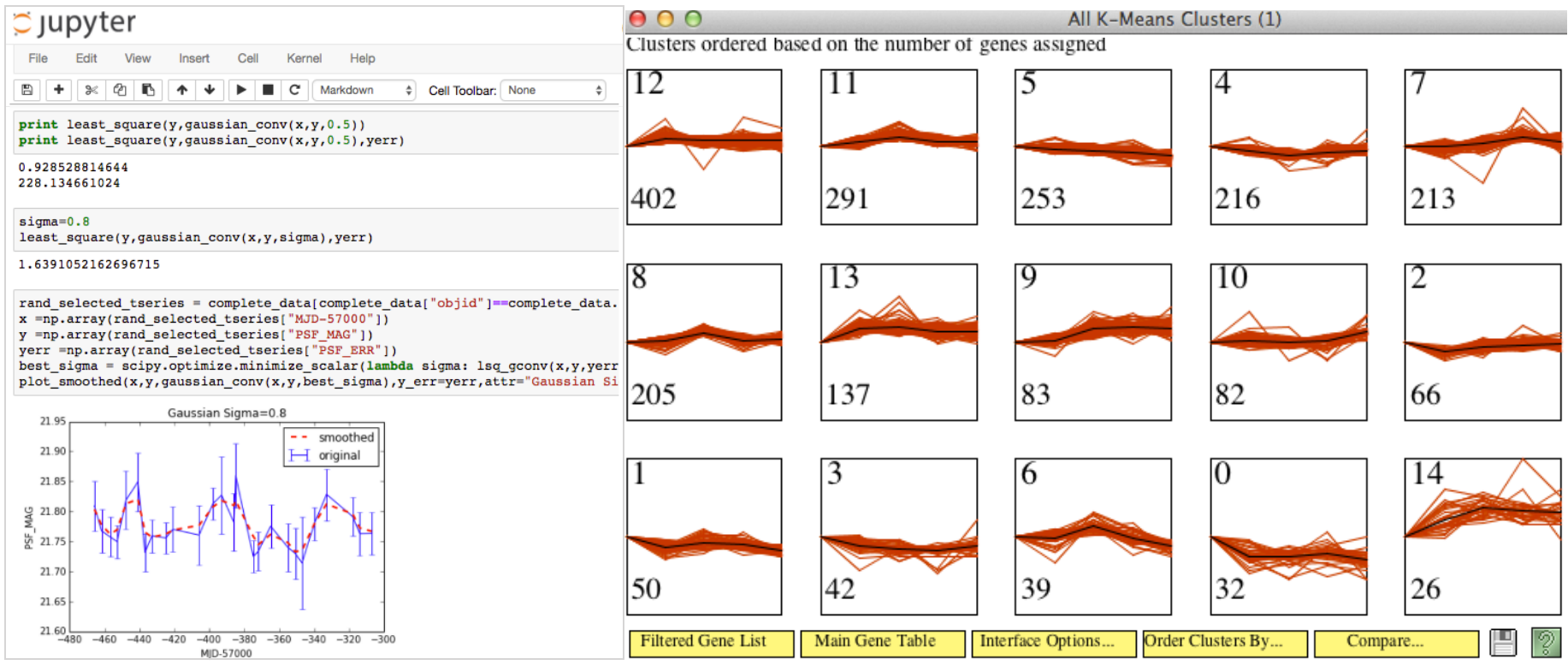}
	   \caption{\rchange{Screenshots from contextual inquiry. Left: A1 performs data smoothing to clean the data and then examines a light curve manually using a Jupyter notebook. Right: G2 uses a domain-specific software to perform clustering and visualize the outputs.}}
	   \label{CIscreenshot}
	   \vspace*{-15pt}
	 \end{figure}
 }
 \par\sectitle{Existing Workflow and Design Opportunities:} 
 \npar \rchange{Since astronomical datasets are often terabytes in scale, they are often processed and stored in highly specialized data management systems in supercomputing centers. As a preliminary step, the astronomer downloads a data sample to explore in a Jupyter notebook, performs data cleaning and wrangling, and verifies data fidelity by computing a set of relevant statistics. Then, to identify transients, the primary scientific goal of their exploration,} the astronomer programmatically \rchange{generates} visualizations of candidate objects with \texttt{matplotlib} and visually \rchange{examines} each light curve. \cut{\rchange{Figure~\ref{CIscreenshot} (left) shows an example of such a manually-generated light curve.}} If an object of interest \rchange{is} identified through visual analysis, the astronomer may inspect the image of the object for verifying that the significant change in brightness was not due to an imaging artifact. 
While experienced astronomers \rchange{like A1 who have} examined many transient light curves can often distinguish an interesting transient from noise by sight, manual searching for transients is still very time-consuming and error-prone, since the large majority of objects are false-positives. \rchange{A1 immediately recognized the potential of VQSs, since he could use} specific pattern \rchange{search} queries to directly identify these rare transients \rchange{ without cumbersome manual examination}.
 \subsection{Genetics}
 \par\sectitle{Participants and Goals:} 
 \npar Gene expression is a common measurement in genetics obtained via microarray experiments~\cite{Peng2016}. We worked with a graduate student (G1) and professor (G3) at a research university who were using gene expression data to understand how genes are related to phenotypes expressed during early embryonic development\techreport{\cite{Peng2016,Gloss2017}}. Their data consisted of a collection of gene expression profiles over time for mouse stem cells, aggregated over multiple experiments. \rchange{Their scientific goal was to \emph{correlate gene function with their expression profiles (i.e., line charts)} by \emph{gaining a high-level overview of the expression profile patterns}.}
 \par\sectitle{Existing Workflow and Design Opportunities:} 
 \npar \rchange{G1 often downloads the raw microarray data from a public database and preprocesses the data using a script written in R. Then, to explore this data, G1 loads the preprocessed gene expression data into a custom desktop application to visualize and cluster the gene expression profiles\cut{, as shown in Figure~\ref{CIscreenshot} (right)}.} Prior to the study, G1 and G3 spent over a month \rchange{searching for the ``right'' number of groups to cluster the profiles, by iteratively tuning the parameters on the clustering application and evaluating the output via a mix of application-provided visualizations and programmatically-generated statistics.}\cut{attempting to determine the best number of clusters for grouping together these gene expression profiles based on a series of statically-generated visualizations and statistics computed after clustering.} While regenerating their results took no more than 15 minutes every time they made a change, the multi-step, segmented workflow meant that all changes had to be done offline\rchange{, this is, they could only test out a few variations per week}.\techreport{, so that valuable meeting time was not wasted trying to regenerate results.} \rchange{When we first demonstrated the capabilities of a VQS in our introductory meeting, G3 was astonished to see that on performing an interaction, the recommended visualizations updated almost instantaneously, as opposed to waiting until the next meeting for G1 to re-generate the results. They expressed an interest in VQSs, since the tool had the potential to dramatically speed up their collaborative analysis process.}
 \subsection{Material Science}
 \par\sectitle{Participants and Goals:} 
 \npar We collaborated with material scientists at a research university who identify solvents for energy-efficient and safe batteries. These scientists worked on a large simulation dataset containing chemical properties for more than 280,000 solvents~\cite{Khetan2018}. Each row of their dataset corresponded to a unique solvent with 25 different chemical attributes. We worked closely with a postdoctoral researcher (M1), professor (M2), and graduate student (M3) to design a sensible way of exploring their data. They wanted to use VQSs to discover solvents that not only have similar properties to known solvents, but are also more favorable (e.g., cheaper or safer to manufacture). To search for these solvents, they needed to {\em\rchange{understand how changes in certain chemical properties affect others (expressed as trends in line charts) under specific conditions.}}
 \par\sectitle{Existing Workflow and Design Opportunities:} 
 \npar M1 typically \rchange{starts} his data exploration process by applying filters to a list of potential battery solvents using SQL queries \rchange{(e.g., find solvents with boiling point over 300 Kelvins and lithium solvation energy under 10 kcal/mol)}. \rchange{By iteratively applying and adjusting different (often complementary) sets of filters, he compares between different groups of solvents by observing their properties across a small sample. He manually examines the properties of each individual solvent by inspecting} the 3D chemical structure of the solvent in a custom software, as well as gathering information regarding the solvent by cross-referencing an external chemical database and existing uses of this solvent in literature. The collected information, including cost, availability, and other physical properties, enabled researchers to select the final set of desirable solvents that could be feasibly experimented with in their lab. \rchange{While M1 could identify potential solvents through manual lookups and comparisons, M2 and M1 saw the value in VQSs since it was often impossible to manually uncover hidden relationships between different attributes, such as how changes in one property affects the behavior of others for a class of solvents, across large numbers of solvents.} 
 \subsection{\rchange{Themes Emerging From Need-finding Phase}}
 \par \rchange{Across the domains, several themes emerged around the bottlenecks that participants experienced in existing workflows.
 \begin{denselist}
	 \item \textbf{Need for Expressive Querying:} While there is often a need to compare among large numbers of data instances, it is difficult to express and search for a desired shape-based pattern through programming languages like SQL or Python. And yet, none of the participants have heard of VQSs, let alone use them.
	 \item \textbf{Need for Integrative Workflows:} Users often switched between different analytical tasks, including preprocessing, parameter specification, code execution, and visualization comparisons. The non-interactive nature of their segmented workflows impedes exploratory analysis and hinders collaboration.
	 \item \textbf{Need for Faceted Exploration:} To deal with the large volume of data present, users have to select particular samples or subsets of data that are ``worth investigating''. Often, the choice of what criteria to apply as filters is also exploratory.
\end{denselist}
 These themes seeded the collaborative feature discovery process, leading to the development of the system prototype, described next}.
 
\begin{table*}[ht!]
\centering
  \resizebox{0.9\textwidth}{!}{%
\begin{tabular}{|p{0.05cm}|l|l|l|l|l|}
	\hline
	                                                    & Component                                                                                                 & Feature                                                                                                              & Purpose                                                                                                                                                                                                      & Task Example                                                                                                                                                                                                                                                    & \begin{tabular}[c]{@{}l@{}}Similar Features\\ in Past VQSs\end{tabular}                                                                                                                                                                                         \\ \hline
	\rowcolor[HTML]{AADFFD} 
	\cellcolor[HTML]{AADFFD}                                   & \cellcolor[HTML]{AADFFD}                                                                                  & \begin{tabular}[c]{@{}l@{}}Query by Sketch\\ (Figure \ref{zvOverview}B1)\end{tabular}               & \begin{tabular}[c]{@{}l@{}}Freehand sketching for \\ specifying pattern query.\end{tabular}                                                                                                                  & \begin{tabular}[c]{@{}l@{}}\A Find patterns with a peak \\ and long-tail decay that\\ may be supernovae candidates.\end{tabular}                                                                                                                 & \begin{tabular}[c]{@{}l@{}}All include sketch \\ canvas except~\cite{Hochheiser2004}.\end{tabular}                                                                                                                                        \\ \cline{3-6} 
	\rowcolor[HTML]{AADFFD} 
	\cellcolor[HTML]{AADFFD}                                   & \cellcolor[HTML]{AADFFD}                                                                                  & \begin{tabular}[c]{@{}l@{}}Input Equation\\ (Figure \ref{zvOverview}A1)\end{tabular}                & \begin{tabular}[c]{@{}l@{}}Specify a exact functional \\ form as a pattern query \\ (e.g., y=$x^2$).\end{tabular}                                                                                            & \begin{tabular}[c]{@{}l@{}}\M Find patterns exhibiting \\ inversely proportional \\ chemical relationship.\end{tabular}                                                                                                                          & ----                                                                                                                                                                                                                                                            \\ \cline{3-6} 
	\rowcolor[HTML]{AADFFD} 
	\cellcolor[HTML]{AADFFD}                                   & \multirow{-5}{*}{\cellcolor[HTML]{AADFFD}\begin{tabular}[c]{@{}l@{}}\textbf{Pattern Specification:}\\\textit{What is the shape of}\\\textit{the pattern query?}\end{tabular}} & \begin{tabular}[c]{@{}l@{}}Pattern Upload\\ (Figure \ref{zvOverview}D2)\end{tabular}                & \begin{tabular}[c]{@{}l@{}}Upload a pattern consisting\\ of a sequence of points as \\ a query.\end{tabular}                                                                                                 & \begin{tabular}[c]{@{}l@{}}\A Find supernovae based on \\ previously discovered sources.\end{tabular}                                                                                                                                            & \begin{tabular}[c]{@{}l@{}}Upload CSV\\ \cite{mohebbi2011google}\end{tabular}                                                                                                                                                                  \\ \cline{2-6} 
	\rowcolor[HTML]{AADFFD} 
	\cellcolor[HTML]{AADFFD}                                   & \cellcolor[HTML]{AADFFD}                                                                                  & \begin{tabular}[c]{@{}l@{}}Smoothing\\ (Figure \ref{zvOverview}D2)\end{tabular}                     & \begin{tabular}[c]{@{}l@{}}Interactively adjusting the level \\ of denoising on visualizations,\\ effectively changing the degree\\ of shape approximation when \\ performing pattern matching.\end{tabular} & \begin{tabular}[c]{@{}l@{}}\textbf{A, M:} Eliminate patterns \\ matched to spurious noise.\end{tabular}                                                                                                                                        & \begin{tabular}[c]{@{}l@{}}Smoothing ~\cite{Mannino2018}\\ Angular slope queries ~\cite{Hochheiser2004}\\ Trend querylines ~\cite{ryall2005querylines}\end{tabular}                     \\ \cline{3-6} 
	\rowcolor[HTML]{AADFFD} 
	\cellcolor[HTML]{AADFFD}                                   & \cellcolor[HTML]{AADFFD}                                                                                  & \begin{tabular}[c]{@{}l@{}}Range \\ Selection\\ (Figure \ref{zvOverview}B2, D4)\end{tabular}        & \begin{tabular}[c]{@{}l@{}}Restrict to query only in \\ specific x/y ranges of interest \\ through brushing selected\\ x-range and filtering \\ selected y-range.\end{tabular}                               & \begin{tabular}[c]{@{}l@{}}\A Matching only around \\ shape exhibiting a peak.\\ \M Matching only around \\ shape region that exhibit linear\\ or exponential relationships\end{tabular}                                          & \begin{tabular}[c]{@{}l@{}}Text Entry ~\cite{wattenberg2001sketching,Mannino2018}\\ Min/max boundaries ~\cite{ryall2005querylines}\\ Range Brushing ~\cite{Hochheiser2001}\end{tabular} \\ \cline{3-6} 
	\rowcolor[HTML]{AADFFD} 
	\multirow{-20}{*}{\cellcolor[HTML]{AADFFD}\rot{\vspace{-2pt}Top-Down}}         & \multirow{-10}{*}{\cellcolor[HTML]{AADFFD}\begin{tabular}[c]{@{}l@{}}\textbf{Match Specification:}\\\textit{How should the pattern}\\\textit{query be matched} \\\textit{with other visualizations?}\end{tabular}}   & \begin{tabular}[c]{@{}l@{}}Range \\ Invariance\\ (Figure \ref{zvOverview}D1,4)\end{tabular}         & \begin{tabular}[c]{@{}l@{}}Ignoring vertical or horizontal \\ differences in pattern matching \\ through option for x-range\\ normalization and y-invariant\\ similarity metrics .\end{tabular}              & \begin{tabular}[c]{@{}l@{}}\A Searching for existence of a\\ peak above a certain amplitude.\\ \G Searching for a \\ ``generally-rising" pattern.\end{tabular}                                                                    & \begin{tabular}[c]{@{}l@{}}Temporal invariants ~\cite{correll2016semantics}\end{tabular}                                                                                                                                                \\ \hline
	\rowcolor[HTML]{FBE39C} 
	\cellcolor[HTML]{FBE39C}                                   & \cellcolor[HTML]{FBE39C}                                                                                  & \begin{tabular}[c]{@{}l@{}}Data selection\\ (Figure \ref{zvOverview}A)\end{tabular}                 & \begin{tabular}[c]{@{}l@{}}Changing the collection of \\ visualizations to iterate over.\end{tabular}                                                                                                        & \begin{tabular}[c]{@{}l@{}}\M Explore tradeoffs and \\ relationships between \\ physical attributes.\end{tabular}                                                                                                                                & ----                                                                                                                                                                                                                                                            \\ \cline{3-6} 
	\rowcolor[HTML]{FBE39C} 
	\cellcolor[HTML]{FBE39C}                                   & \multirow{-4}{*}{\cellcolor[HTML]{FBE39C}\begin{tabular}[c]{@{}l@{}}\textbf{View Specification:} \\ \textit{What data to visualize} \\ \textit{and how should it} \\ \textit{be displayed?}\end{tabular}}    & \begin{tabular}[c]{@{}l@{}}Display control\\ (Figure \ref{zvOverview}D4)\end{tabular}               & \begin{tabular}[c]{@{}l@{}}Changing the details of \\ how visualizations should\\ be displayed.\end{tabular}                                                                                                 & \begin{tabular}[c]{@{}l@{}}\M Non-time-series data should \\ be displayed as scatterplot.\end{tabular}                                                                                                                                           & ----                                                                                                                                                                                                                                                            \\ \cline{2-6} 
	\rowcolor[HTML]{FBE39C} 
	\cellcolor[HTML]{FBE39C}                                   & \cellcolor[HTML]{FBE39C}                                                                                  & \begin{tabular}[c]{@{}l@{}}Filter\\ (Figure \ref{zvOverview}D3)\end{tabular}                        & \begin{tabular}[c]{@{}l@{}}Display and query only on data \\ that satisfies the composed \\ filter constraints.\end{tabular}                                                                                 & \begin{tabular}[c]{@{}l@{}}\A Eliminate unlikely \\ candidates by navigating to \\ more probable data regions.\\ \textbf{M, G:} Compare how overall\\ patterns change when filtered \\ to particular data subsets.\end{tabular} & ----                                                                                                                                                                                                                                                            \\ \cline{3-6} 
	\rowcolor[HTML]{FBE39C} 
	\multirow{-10}{*}{\cellcolor[HTML]{FBE39C}\rot{\vspace{-2pt}Context Creation}} & \multirow{-7}{*}{\cellcolor[HTML]{FBE39C}\begin{tabular}[c]{@{}l@{}}\textbf{Slice-and-Dice:} \\ \textit{How does navigating} \\ \textit{to another data subset} \\ \textit{change the query result?}\end{tabular}}                                                  & \begin{tabular}[c]{@{}l@{}}Dynamic Class \\ (Figure~\ref{dcc})\end{tabular}                    & \begin{tabular}[c]{@{}l@{}}Create custom classes of data \\ that satisfies one or more \\ specified range constraints. \\ Display aggregate \\ visualizations for separate\\ data classes.\end{tabular}      & \begin{tabular}[c]{@{}l@{}}\textbf{A, M:} Examine aggregate \\ patterns of different data \\ classes.\end{tabular}                                                                                                                             & ----                                                                                                                                                                                                                                                            \\ \hline
	\rowcolor[HTML]{B5E1A4} 
	\cellcolor[HTML]{B5E1A4}                                   & \begin{tabular}[c]{@{}l@{}}\textbf{Result Querying:} \\ \textit{What other visualizations}\\ \textit{``look similar" to the} \\\textit{selected pattern?}\end{tabular}                                                & \begin{tabular}[c]{@{}l@{}}Drag-and-drop\\ (Figure \ref{zvOverview}C, E)\end{tabular}               & \begin{tabular}[c]{@{}l@{}}Querying with any selected\\ result visualization as pattern\\ query (either from \\ recommendations or results).\end{tabular}                                                    & \begin{tabular}[c]{@{}l@{}}\textbf{A, G, M:} Find other objects that\\ are similar to X; Examine what \\ other objects similar to X look \\ like overall.\end{tabular}                                                                         & \begin{tabular}[c]{@{}l@{}}Drag-and-drop ~\cite{Hochheiser2001}\\ Double-Click ~\cite{correll2016semantics}\end{tabular}                                                                                        \\ \cline{2-6} 
	\rowcolor[HTML]{B5E1A4} 
	\multirow{-6}{*}{\cellcolor[HTML]{B5E1A4}\rot{\vspace{-2pt}Bottom-Up}}        & \begin{tabular}[c]{@{}l@{}}\textbf{Recommendation:} \\ \textit{What are the key patterns} \\ \textit{in this dataset?}\end{tabular} & \begin{tabular}[c]{@{}l@{}}Representative \\ and Outliers\\ (Figure \ref{zvOverview}E)\end{tabular} & \begin{tabular}[c]{@{}l@{}}Displaying visualizations of \\ representative trends and outlier\\ instances based on clustering.\end{tabular}                                                                   & \begin{tabular}[c]{@{}l@{}}\A Examine anomalies and debug \\ data errors through outliers.\\ \textbf{G, M:} Understand representative \\ trends common to this dataset \\ (or filtered subset).\end{tabular}                    & ----                                                                                                                                                                                                                                                            \\ \hline
\end{tabular}
}
  \caption{\rchange{Taxonomy of key capabilities essential to VQSs and major features incorporated via user-centered design. We organize each feature based on its functional component. From left to right, each of the three sensemaking processes (first column) is broken down into key functional components (second column) in VQSs. Each component addresses a pro-forma question from the system's perspective.} Table cells are further colored according to the sensemaking process that each component corresponds to (Blue: Top-down, Yellow: Context creation, Green: Bottom-up). We list the functional purpose of each feature based on how it is implemented in \zvpp, example use cases from participatory design (\A astronomy, \M material science, \G genetics), and similar features incorporated in past VQSs. Given the exhaustive nature of Table~\ref{bigfeaturetable}, each motivated by example use cases from one or more domains, we further organize the features in terms of the Section~\ref{sec:sensemaking} sensemaking framework and assess their effectiveness in the Section~\ref{sec:eval_findings} evaluation study.}\label{bigfeaturetable}
  \vspace*{-15pt}
\end{table*}
 \section{\rchange{Design} Process and System Overview\label{sec:pd_findings}}
 Given the need for a VQS, we further collaborated with participants to develop features to address their problems and challenges \rchange{in Phase II of our study}. We \rchange{first provide a high-level system overview of the \rchange{design} product, \zvpp, then we reflect on our feature discovery process}.
 \subsection{System Overview\label{sec:system}}
  The \zvpp interface is organized into 5 major regions all of which dynamically update upon user interactions. Typically, participants begin their analysis by selecting the dataset and attributes to visualize in the \emph{data selection panel} (Figure~\ref{zvOverview}A). Then, they specify a pattern of interest as a query (hereafter referred to as \emph{pattern query}), through either sketching, inputting an equation, uploading a data pattern, or dragging and dropping an existing visualization, displayed on the \emph{query canvas} (Figure~\ref{zvOverview}B). \zvpp performs shape-matching between the queried pattern and other possible visualizations, and returns a ranked list of visualizations that are most similar to the queried pattern, displayed in the \emph{results panel} (Figure~\ref{zvOverview}C). At any point during the analysis, analysts can adjust various system-level settings through the \emph{control panel} (Figure~\ref{zvOverview}D) or browse through the list of \emph{recommendations} provided by \zvpp (Figure~\ref{zvOverview}E). For comparison, the existing \zv system \cut{(Figure~\ref{oldZV} in Appendix~\ref{apdx:pdartifact}) }from~\cite{Siddiqui2017} allowed users to query via sketching or drag-and-drop and displayed representative and outlier pattern recommendations, but had limited capabilities to navigate across different data subsets and had few control settings. Our \zvpp system is open source and available at: \url{http://github.com/zenvisage/zenvisage}; \rchange{other details and documentation can be found at that link}. 
 \subsection{The Collaborative Feature Discovery Process~\label{sec:feature_dsicovery}}
 \par Throughout the \rchange{design} process, we worked closely with participants to discover VQS capabilities that were essential for addressing their high-level domain challenges. We identified various subtasks based on the participant's workflows, designed sensible features for accomplishing these subtasks that could be used in conjunction with existing VQS capabilities, and elicited feedback on intermediate feature prototypes. Bodker et al.~\cite{BodkerGronbaek} cite the importance of encouraging user participation and creativity in cooperative design through different techniques, such as future workshops, critiques, and situational role-playing. Similarly, our objective was to collect as many feature proposals as possible\cut{, while being inclusive across different domains}. We further organized these features \rchange{we added to \zvpp} into Table~\ref{bigfeaturetable} through an iterative coding process~\cite{Muller2012} by one of the authors.
 \par \cut{In grounded theory methods~\cite{Muller2012}, researchers first create \emph{open codes} to assign descriptive labels to raw data, followed by grouping open codes together by relationships or categories to form \emph{axial codes}. Finally, \emph{selective codes} are obtained by focusing on specific sets of axial codes to come up with a set of core emerging concepts. Inspired by grounded theory methods,} We first collected the list of features\rchange{, }example usage scenarios\rchange{, }and similar capabilities in existing VQSs as open codes\rchange{, corresponding to individual rows in Table~\ref{bigfeaturetable}}. Then, we further organized this list into axial codes representing ``components''\rchange{: core functionalities essential to VQSs (second} column in Table~\ref{bigfeaturetable}). Finally, \cut{as we will describe in Section~\ref{sec:sensemaking}, }the selective codes capture each of the sensemaking processes (\rchange{leftmost column} in Table~\ref{bigfeaturetable}). Instead of describing this table in detail, we present a typical example of how this table is organized. \rchange{From right to left, consider the row corresponding to the Smoothing feature (column 3) in} Table~\ref{bigfeaturetable}: one of the common challenges in astronomy and material science is that noise in the dataset can result in large numbers of false-positive matches. To address this issue, smoothing is a feature in \zvpp that enables users to adjust data smoothing algorithms and parameters on-the-fly to both denoise the data and change the degree of shape approximation applied when performing pattern matching. 
 Smoothing, along with range selection and range invariance\cut{(row 5 and 6)}, is part of the \emph{match specification} component: VQS mechanisms for clarifying how matching should be performed. Both match specification and \emph{pattern specification} (a description of what the pattern query should look like) are essential components for supporting the sensemaking process top-down pattern search (in blue\rchange{, as labeled in the leftmost column}).
 \cut{\begin{figure}[h!]
 	\centering
   \includegraphics[width=\linewidth]{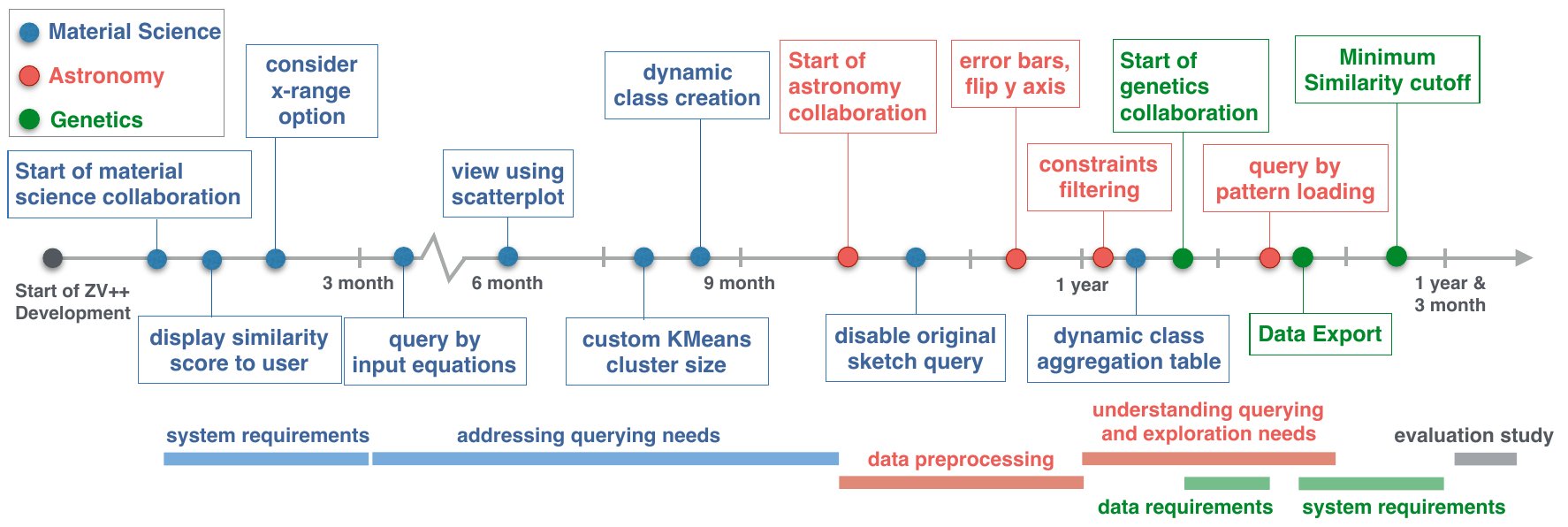}
 	\caption{Timeline for progress in participatory design studies.}
 	\label{timeline}
 \end{figure}}
 \par It is important to note that while some of the proposed features in Table~\ref{bigfeaturetable} (such as data filtering and view specification) are pervasive in general visual analytics (VA) systems~\cite{Heer2012,Amar2005}, they have not been incorporated in present-day VQSs. In fact, one of the key insights here is in recognizing the need for an \emph{integrative} VQS whose sum is greater than its parts, that encourages analysts to rapidly generate hypotheses and discover insights by facilitating all three sensemaking processes. This finding is partially enabled by the unexpected benefits that come with collaborating with multiple groups of participants during the feature discovery process. \rchange{Next, we reflect on what worked and what didn't work in the feature discovery process, to inform similar design studies for visual analytics systems.
 %
 }
 \rchange{
 \par \sectitle{Cross-pollination and Generalization via Parallel Use Cases.}}
 \npar Introducing the newly-added features to \zvpp that addressed a particular domain often resulted in unexpected use cases for other domains. Considering feature proposals from multiple domains can also \rchange{result in cross-pollination of feature designs, often leading to} more generalized design choices. For example, around the same time when we spoke to astronomers who wanted to eliminate sparse time series from their search results, our material science collaborators also expressed a need for inspecting only solvents with properties above a certain threshold. \rchange{Instead of developing separate domain-specific features,} data filtering arose as a crucial, common operation that was later incorporated into \zvpp to support this class of queries. 
 \rchange{
 \par \sectitle{The Hidden Upfront Cost of Domain Integration.}
 \npar While we expected to spend most of our collaborative design effort on figuring out the mechanics of visual query specification and matching, instead, preparing participant datasets for use in our system by meeting data and system requirements was the most time-consuming aspect of this phase\footnote{We provide a detailed timeline in Appendix~\ref{apdx:pdartifact}.}. Data requirements include gaining an understanding of the problem domain, understanding the types of data suitable for a VQS, and cleaning and loading of this data. System requirements include features required for the data to be visualized appropriately. Often, participants could only envision the types of queries to issue and how variations to the system could help better address their needs after seeing their data displayed for the first time in the prototype. 
 We also found that the time it took us to satisfy the data and system requirements decreased as we progressed to the later domains, by leveraging existing features in our prototype to satisfy some of
 the upfront needs. 
 \cut{In summary, it is important to allocate sufficient time when working with real-world datasets and users to account for initial system and data challenges, as well as developing general-purposed integration features (such as data uploading tools) that could be used across multiple use cases to decrease the upfront cost for future collaborations with new domains.}
 \par \sectitle{Build Connectors, not Swiss-Army Knives.} 
 \npar Participants often envisioned how VQSs can be used in conjunction with other resources that they are familiar with, including those used for reference, computing statistics, browsing related datasets, or examining other data attributes or visualization types not supported in the VQS (scatterplots, histograms). The prevalence of external tools for supporting analytical inquiries stems from how analysts often require multiple data sources or data attributes to further develop or verify their hypothesis. For example, to determine whether a particular gene belongs to a regulatory network, G2 not only needed to look at the expression data in the VQS, but also enrichment testing and knockout data. Likewise, others used specialized tools for visualizing telescope images and 3D chemical structures. Instead of forcing our VQS prototype into a swiss-army knife, we instead focused on building connectors that enable smoother transitions between tools. For example, our data upload and pattern upload feature invites participants to bring data from an external tool into \zvpp, while our data export feature allowed users to download the similarity, representative trend, and outlier results as csv files from \zvpp into an external tool\cut{for downstream analysis, or export individual visualizations to facilitate easier sharing of visualization results with collaborators}. For example, geneticists could export the clusters directly from \zvpp as inputs to their downstream regression analysis.
 \par \sectitle{The Art of Problem Selection.}}
 \npar While our collective brainstorming led to the cross-pollination and generalization of features, this technique can also lead to unnecessary features that result in wasted engineering effort. \rchange{During co-design}, there were numerous \cut{problems and associated }features proposed by participants, not all of which were incorporated. \rchange{The reasons for not carrying a feature from design to implementation stage included:
 \begin{denselist} 
 \item Nice-to-haves: One of the most common reasons for unincorporated features comes from participant's requests for nice-to-have features. We use two criteria (necessity and generality across domains) to judge whether to invest in developing a particular feature.
 \item ``One-shot'' operations: We decided not to include features that only needed to be performed once and remain fixed thereafter in the analysis workflow. For example, certain preprocessing operations such as filtering null values only needed to be performed once with an external tool, whereas data smoothing is a procedure that requires some degree of tuning and adjustments.
 \item Substantial research or engineering effort: Some proposed features did not make sense in the context of VQS or required a completely different set of research questions. For example, the question of how to properly compute similarity between time series with non-uniform number of datapoints arose in the astronomy and genetics use case, but requires the development of a novel distance metric and algorithm that is out of the scope of our design study objective. 
 \item Underdeveloped ideas: Other feature requirements came from casual specification that was underspecified. For example, A1 wanted to look for objects that have a deficiency in one band and high emission in another band, but the scientific definition of ``deficiency'' in terms of brightness levels was ambiguous.
 \end{denselist}
 \npar The decision of whether to invest in developing a feature requires a careful balance between promoting unforseen feature and wasted engineering efforts. Failure to identify these early signs may result in feature implementations that turn out not to be useful for the participants or result in feature bloat.
 } 
 
 \section{A Sensemaking Model for VQSs\label{sec:sensemaking}}
 We now revisit Table~\ref{bigfeaturetable} in an effort to contextualize our \rchange{design} findings using Pirolli and Card's sensemaking framework~\cite{Pirolli}. 
 \rchange{Pirolli and Card's sensemaking model for expert intelligence analysis distinguishes between information processing tasks that are \textit{top-down} (from theory to data) and \textit{bottom-up} (from data to theory). Correspondingly, in the context of VQSs,} analysts can query either directly based on a pattern ``in their head''~\cite{Sedlmair2012} via \emph{top-down pattern specification} or based on the data or visualizations presented to them by the system via \emph{bottom-up data-driven inquiry}. In addition, when analysts do not know what attributes to visualize, \emph{context creation} helps analysts navigate across different collections of visualizations to seek visualization attributes of interest. \cut{A more detailed articulation of the problem space addressable by VQS and how each sensemaking process fits into this space can be found in Appendix~\ref{appdx:problem_space}.}In this section, we first describe \rchange{the objectives of each sensemaking process, then we discuss} how each sensemaking process is comprised of functional components that address the problem and dataset characteristics of each domain. \cut{For reference, the mapping between specific \zvpp features and these components and processes can be found in \rchange{the left two columns of} Table~\ref{bigfeaturetable}.}
   \subsection{Top-Down Pattern Search}
   Top-down processes are ``\textit{goal-oriented}'' tasks that make use of ``\textit{analysis or re-evaluation of theories [and] hypotheses [to] generate new searches}''~\cite{Pirolli}. Applying this notion to the context of VQSs, the goal of top-down pattern search is to search for data instances that exhibit a specified pattern, based on analyst's intuition about how the desired patterns should look like ``in theory'' (including visualizations from past experience or abstract conceptions based on external knowledge). Based on this preconceived notion of what patterns to search for, the design challenge is to translate the pattern query from the analyst's head to a query executable by the VQS. This requires both components for specifying the pattern (\textit{pattern specification}), as well as controls governing how the pattern-matching is performed (\textit{match specification}).
   \boldpara{Pattern Specification} interfaces allow users to submit exact descriptions of a pattern query. This is useful when the dataset contains \emph{large numbers of potentially-relevant pattern instances}.
   Since it is often difficult to sketch precisely, additional shape characteristics of the pattern query (e.g., patterns containing a peak with a known amplitude, or expressible as a functional form) can be used to further winnow the list of undesired matches.
   \boldpara{Match Specification} addresses the well-known problem in VQSs where pattern queries are imprecise~\cite{correll2016semantics,Holz2009,Eichmann2015} by enabling users to clarify how pattern matching should be performed. Match specification is useful when the dataset is \emph{noisy}. When the pattern query satisfies some additional constraints (e.g., the pattern is \rchange{horizontally} invariant), adjusting these knobs \cut{helps }prune away matches that are false-positives to help analysts discover true desired candidates.
   \boldpara{Usage Scenario:} A1 knows intuitively what a supernovae pattern should look like and its detailed shape characteristics, such as the amplitude of the peak and the level of error tolerance for defining a match. He \rchange{first} performs top-down pattern search by querying for transient patterns through sketching\rchange{, then adjusts} the match criterion by choosing to ignore differences along the temporal dimension and changing the similarity metric for flexible matching.
   \subsection{Bottom-Up Data-Driven Inquiry}
   In Pirolli and Card's sensemaking model, bottom-up processes are ``\textit{data-driven}'' tasks initiated by ``\textit{noticing something of interest in data}''~\cite{Pirolli}. Likewise in VQSs, bottom-up data-driven inquiry is a browsing-oriented sensemaking process that involves tasks that are inspired by system-generated visualizations or results. The design challenge for VQSs to support bottom-up inquiries is to develop the right set of ``stimuli'' through \textit{recommendations} that could provoke further data-driven inquiries, as well as low-effort mechanisms to search via these pattern instances through \textit{result querying}. As we will discuss later, this process is crucial but underexplored in past work on VQSs. 
   \boldpara{Recommendations} display visualizations that may be of interest to users based on the current data context. In \zvpp, recommendations comprise of representative trends and outliers, which are useful for understanding common and outlying behaviors when a \emph{small number of common patterns} is exhibited in the dataset. 
   \boldpara{Result querying} enables users to query for patterns similar to a selected data pattern from the ranked list of results or recommendations. Typically, analysts select visualizations with \emph{semantic or visual properties} of interest and make use of result querying to understand characteristic properties of similar instances.
   \boldpara{Usage Scenario:} G2 does not have an upfront knowledge of what to search for. She learns about the characteristic patterns that exist in the dataset through the representative trends, a form of bottom-up inquiry, as a means to jump-start further queries via result querying, as well as understand groups of data instances with shared characteristics.
   \subsection{Context creation}
   While top-down and bottom-up processes operate on a collection of visualizations with fixed X and Y attributes, context creation operates in the regime where the analyst may be investigating the relationships between multiple different attributes or values of interest. Context creation enables analysts to navigate across different visualization collections to learn about patterns in different regions of the data. The design challenge of context creation is to help users visualize and compare how data changes between these different contexts by constructing visualization collections with different visual encodings (\textit{view specification}) or different data subsets (\textit{slice-and-dice}).
   \boldpara{View specification} settings alter the encoding for all of the visualizations on the VQS currently being examined. This ability to work with different collections of visualizations is useful when the dataset is \emph{multidimensional} and the axes of interest are \emph{unknown}. Modifying the view specification offers analysts different perspectives on the data to locate visualization collections of interest.
   \boldpara{Slice-and-Dice} empowers users to navigate and compare collections of visualizations constructed from different subsets of the data. Data navigation capabilities are essential when the dataset has \emph{large numbers of ``support attributes''} that may be related to the visualization attributes (e.g., geographical location may influence the time series pattern for housing prices). Analysts can either make use of pre-existing knowledge regarding these support attributes to navigate to a data region that is more likely to contain the desired pattern (e.g., filtering to suburbs to find cheaper housing) or discover unknown patterns and relationships between different data subsets (e.g., housing prices are lower in winter than compared to summer).
   \boldpara{Usage Scenario:} M1 recognizes salient trends in his dataset such as inverse or linear correlations, but does not have fixed attributes that he wants to visualize or a pattern in mind to query with. Given a list of physical properties of potential interest, he performs context creation by switching between different visualized attributes to understand the dataset from alternative perspectives. He can also dynamically create different classes of data (e.g., solvents with low solubility or have high capacity) to examine their aggregate patterns.
   \par The three aforementioned sensemaking processes are akin to the well-studied sensemaking paradigms of search (top-down), browse (bottom-up), and faceted navigation (context creation) on the Web~\cite{Hearst2009,Olston2003}. Due to each of their advantages and limitations given different information seeking tasks, search interfaces have been designed to support all three complementary acts and transition smoothly between them to combine the strength of all three sensemaking processes. 
   \cut{Similarly for VQSs, our design objective is to enable all three sensemaking processes in \zvpp. }Our \cut{Section~\ref{sec:eval_findings} }evaluation study reveals that this integrative approach not only accelerates the process of visualization discovery, but also encourages hypotheses generation and experimentation.
 \section{Evaluation Study Findings\label{sec:eval_findings}}
 Based on audio, video screen capture,
 and click-stream logs recorded
 during our Phase III evaluation study,
 we performed thematic analysis via open coding to label every event with a descriptive code. Event codes included specific feature usage,
 insights,
 provoked actions, confusion,
 need for capabilities unaddressed
 by the system, and use of external tools\rchange{\footnote{See Appendix~\ref{apdx:studydetails} for details on our coding protocol.}}. To characterize the usefulness
 of each feature, we further labeled whether each
 feature was useful to a particular participant's analysis.
 A feature was deemed \textit{useful}
 if it was either used in a sensible
 and meaningful way to accomplish a task or address a question during the study,
 or has envisioned usage outside of the constrained
 time limit during the study
 (e.g., if data was available or downstream analysis was conducted).
 \cut{We derived these labels from the study transcript
 to circumvent self-reporting bias~\cite{Williams2017},
 which can often artificially inflate
 the usefulness of the feature under examination.}
 In this section, we will apply our thematic analysis results to understand how each sensemaking process occurs in practice.
 \subsection{Uncovering the Myth of Sketch-to-Insight}
 \par To understand the usefulness of different visual querying modalities, we analyzed their frequency of use in our evaluation study. To our surprise,
 despite the prevalence of sketch-to-query
 systems in the literature, \techreport{Figure \ref{fig:feature_heatmap} shows that} only two out of our nine participants
 found it useful to directly
 sketch a desired pattern onto the canvas. 
 The reason why most participants
 did not find direct sketching useful was that
 they often do not start their analysis with a specific pattern in mind.
 Instead, their intuition about what to query is derived
 from other visualizations they encountered
 during exploration, in which case it makes
 more sense to query using those visualizations
 as examples directly (e.g., by dragging and dropping
 that visualization onto the canvas to submit the query).
 Even if a user has a pattern in mind,
 translating that pattern into a sketch is often hard
 to do. For example,
 A2 wanted to search for a highly-varying signal
 enveloped by a sinusoidal pattern indicating
 planetary rotation \includegraphics[width=3.5\baselineskip,keepaspectratio]{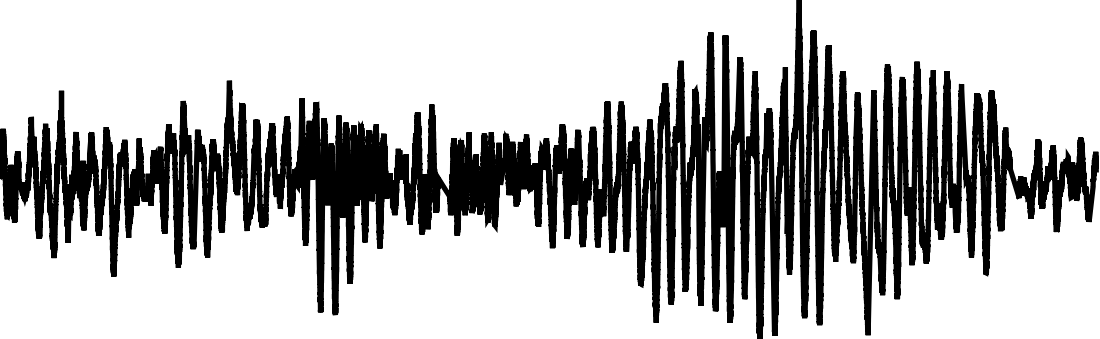}, which was hard to draw by hand.
 \par
\begin{figure}
  \begin{minipage}[b]{.35\linewidth}
     \centering
     \includegraphics[width=\linewidth]{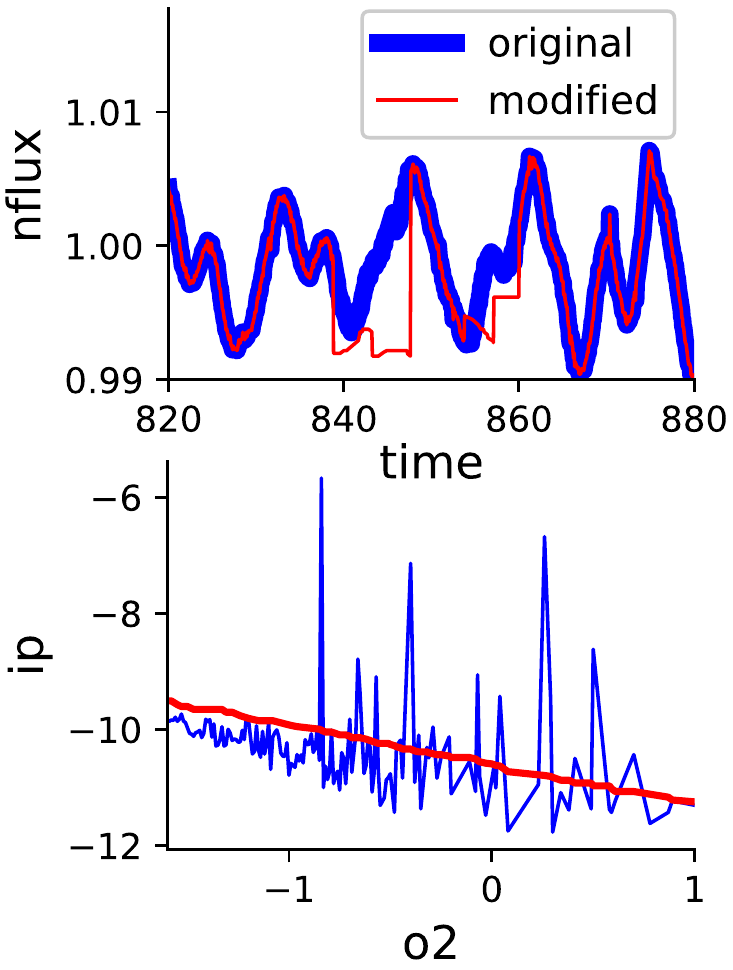}
     \caption{Example of sketch-to-modify, based on canvas traces from M2 (left) and A3 (right). The original drag-and-dropped query is shown in blue and sketch-modified queries in red.}
     \label{query_modification}
     \vspace*{-5pt}
  \end{minipage}
  \hspace{2pt}
  \begin{minipage}[b]{.63\linewidth}
    \centering
    \includegraphics[width=0.9\linewidth]{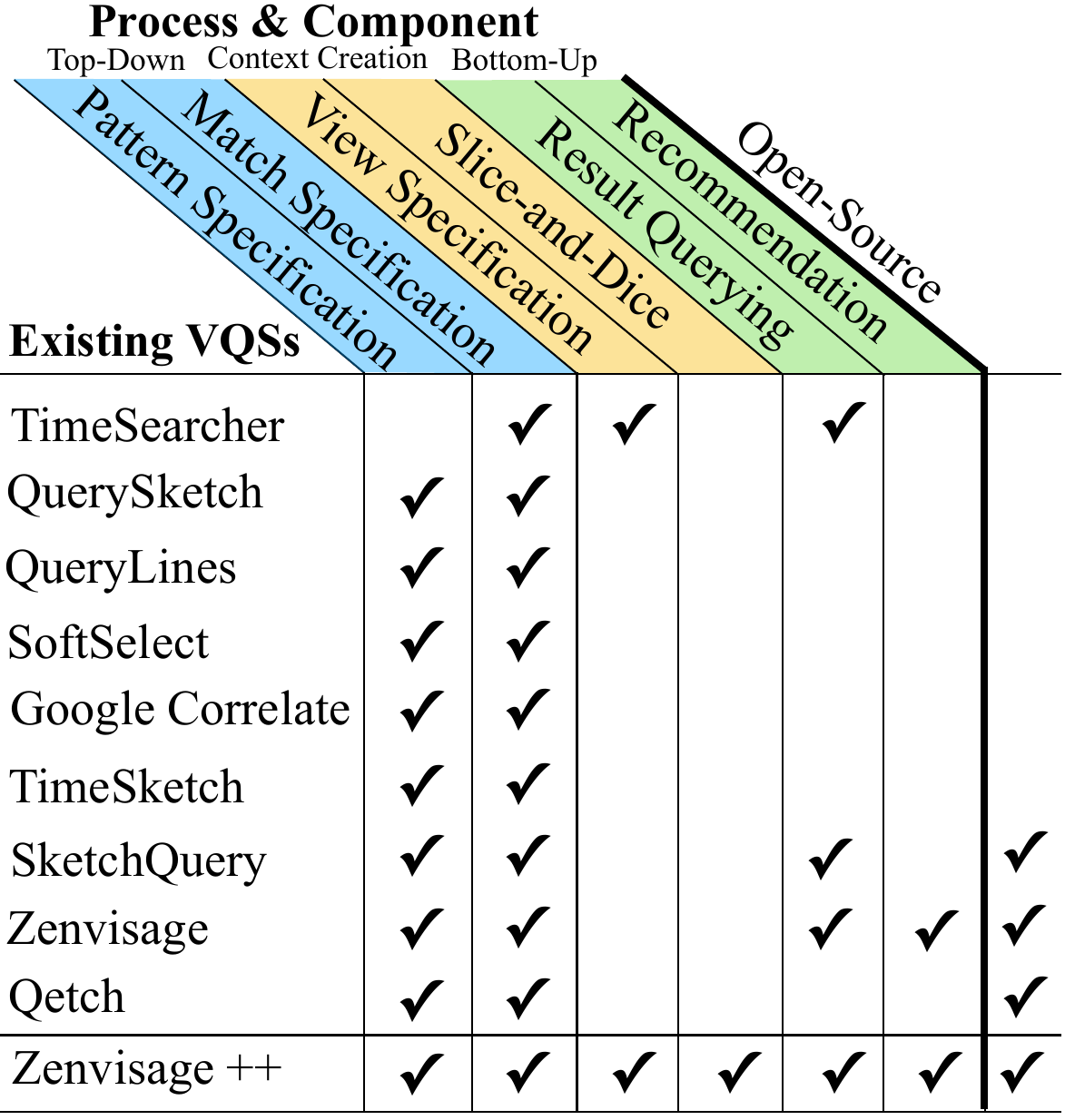}
     \captionof{table}{\cchange{Table summarizing whether key functional components (columns) are covered by past systems (row, ordered by recency)\ccut{, indicated by checked cells}. Column header colors blue, yellow, green represent the three sensemaking processes\ccut{(top-down querying, context creation, and bottom-up querying) described in Section~\ref{sec:pd_findings}}. Heavily-used features for context-creation and bottom-up inquiry are largely missing from prior VQSs.}}
     \label{table:relatedwork}
     \vspace*{-5pt}
  \end{minipage}
  \vspace*{-25pt}
\end{figure}

 We further investigated the processes that participants engaged in to construct pattern queries.\cut{, as presented in Figure~\ref{fig:origins_of_sketch}.} Pattern queries can be generated by either top-down (sketching \rchange{based on user's in-the-head pattern}) or bottom-up (drag-and-drop \rchange{based on what user observes from data}) processes\cut{, driven by various different querying intentions}. \rchange{While our study is not intended as a quantitative study with different querying modalities as conditions, we wanted to get an estimate of the relative frequency of different mechanisms across users. We examined the sequence of interactions that led to each pattern query and labeled each one based on one of the five ways it can be generated---two top-down and three bottom-up ways\footnote{Top-down: sketch-to-query, sketch-to-modify; Bottom-up: Result querying via object of interest, via ranked result, or via recommendations. See Appendix Figure~\ref{fig:origins_of_sketch} for more details.}.}\cut{Figure~\ref{fig:origins_of_sketch} shows that}\rchange{We find that
 \emph{bottom-up processes are 40\% more commonly used than top-down processes for generating a pattern query}}.
  \cut{We will describe the different ways in which the pattern queries are generated \cut{(corresponding to individual bars in Figure~\ref{fig:origins_of_sketch})} in this subsection and the next.}
  Within top-down processes, a pattern query could arise from users directly sketching a new pattern or by modifying an existing sketch. For example, M2 first sketched a pattern to find solvent classes with anticorrelated properties (pattern as a straight line with negative slope) without much success in finding a desired match. So he instead dragged and dropped one
 of the peripheral visualizations similar
 to his desired one and then smoothed
 out the noise in the visualization via sketching, yielding
 a straight line,
 as shown in Figure \ref{query_modification} (left).
 M2 repeated this workflow twice in separate
 occurrences during the study and
 was able to derive insights.
 \rchange{Likewise, A3 was searching for pulsating stars characterized by dramatic changes in the amplitudes of the light curves. She knows that stellar hotspots also exhibit dramatic amplitude fluctuations, but unlike pulsating stars, the variations happen at regular intervals. Figure~\ref{query_modification} (right) illustrates how A3 first picked out a regular pattern (suspected starspot), then modified it slightly so that the pattern looks more \rchange{``irregular''} (to find pulsating stars).}
 \par The infrequent use of top-down pattern
 specification was also reflected in the fact
 that none of the participants queried using an equation.
 In both astronomy and genetics, the visualization patterns
 resulted from complex physical processes
 that could not be written down as equations analytically.
 Even in the case of material science when analytical
 relationships do exist, it is challenging to formulate patterns as functional forms in a prescriptive manner.
 \rchange{\par We found that some users employed match specification to remedy undesired results from their top-down pattern queries. While we did not rigorously study the effects of different analytical parameter settings, we observed that more users refined their matches by adjusting the range and degree of approximation, rather than opting for a different similarity metric. This points to future work in developing more flexible and intuitive vocabularies for modifying the match along the research directions pursued in~\cite{correll2016semantics,Mannino2018} over incorporating additional complex, off-the-shelf matching objectives in VQSs.}
 \par Our findings suggest that while sketching
 is a useful construct for people to express their queries,
 \emph{the existing ad-hoc, sketch-only model for VQSs
 is insufficient on its own} without data examples
 that can help analysts jumpstart their exploration.
 In fact, \cut{from Figure~\ref{fig:origins_of_sketch},
 we can see that sketch-to-query was only used
 8 times, while the remaining querying modalities were used 29 times altogether,
 more than three times as much as sketch-to-query.}\rchange{ we found that sketch-to-query only accounted for about a fifth of the total number of visual queries performed during the study}. This finding has profound implications on the design of future VQSs, since \rchange{our comparison of VQS features across existing work (Table~\ref{table:relatedwork})} suggests that past work has primarily focused on top-down process components, without considering how useful these features are in real-world analytic tasks.
 We suspect that these limitations may be why existing VQSs are not commonly adopted in practice. Note that we are not advocating for removing the natural and intuitive sketch capabilities from future VQSs completely, but instead focusing future research and design efforts to examine other (often underexplored) VQS sensemaking processes. Such processes could be applied in conjunction with sketching to help analysts more flexibly express their analytical goals, described next.

 \subsection{Insights via Context Creation and Bottom-up Approaches}
 \par As alluded to earlier,
 \emph{bottom-up data-driven inquiries
 and context creation are far more commonly
 used than top-down pattern search
 when users have no desired patterns in mind},
 which is typically the case for exploratory data analysis.
 In particular, top-down approaches were only useful for 29\% of the use cases,
 whereas they were useful for 70\% of the use cases
 for bottom-up approaches and 67\%
 for context creation\footnote{See Appendix~\ref{apdx:studydetails} for details on how this measure was computed.}. We now highlight some exemplary workflows demonstrating the efficacy of the latter two sensemaking processes.
 \par \textbf{Bottom-up }pattern queries can come from either
 the ranked list of results,
 recommendations, or by selecting a
 particular object of interest as a drag-and-drop query. \cut{As shown in Figure~\ref{fig:origins_of_sketch} (\bartext{recommendations}),}
 The most common use of bottom-up querying
 is via recommended visualizations. For example, G2 and G3 identified that
 the three representative patterns
 recommended in \zvpp corresponded
 to the same three groups of genes discussed
 in a recent publication~\cite{Gloss2017}:
 induced genes (profiles with expression levels going up \includegraphics[width=2\baselineskip,keepaspectratio]{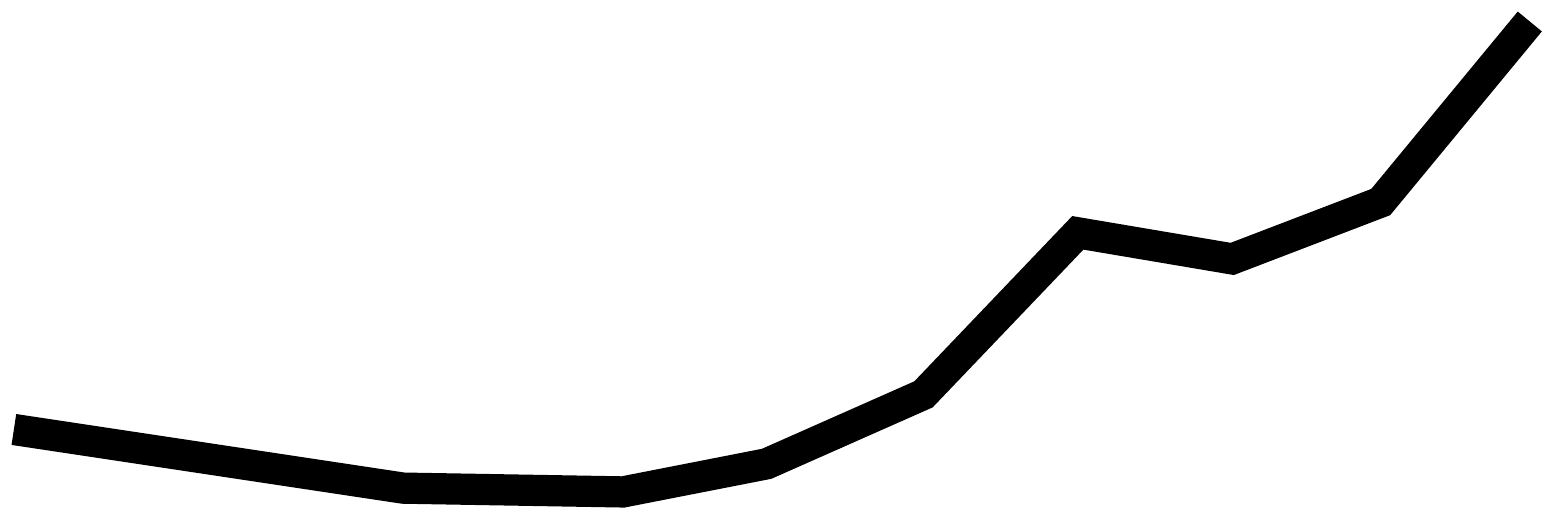}),
 repressed genes (starting high then decreasing \includegraphics[width=2\baselineskip,keepaspectratio]{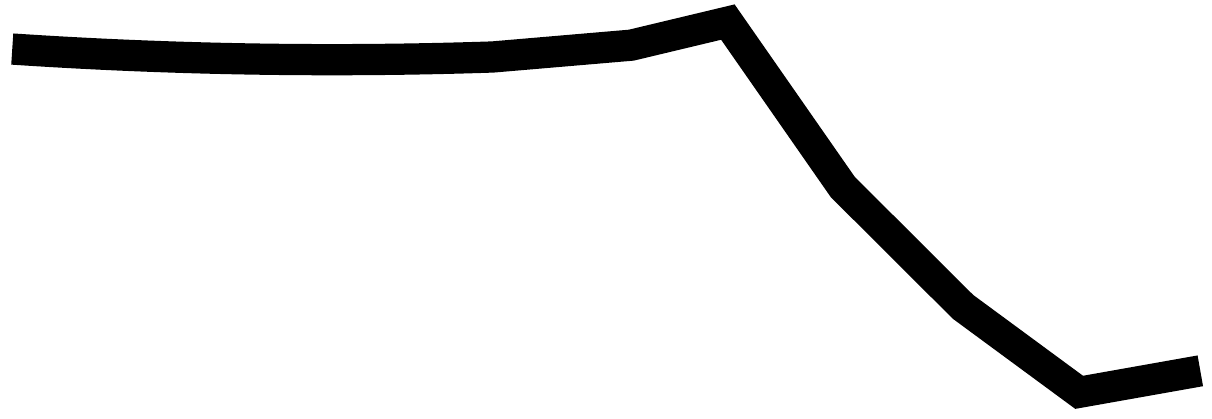}),
 and transients (rising first then dropping at another time point \includegraphics[width=2\baselineskip,keepaspectratio]{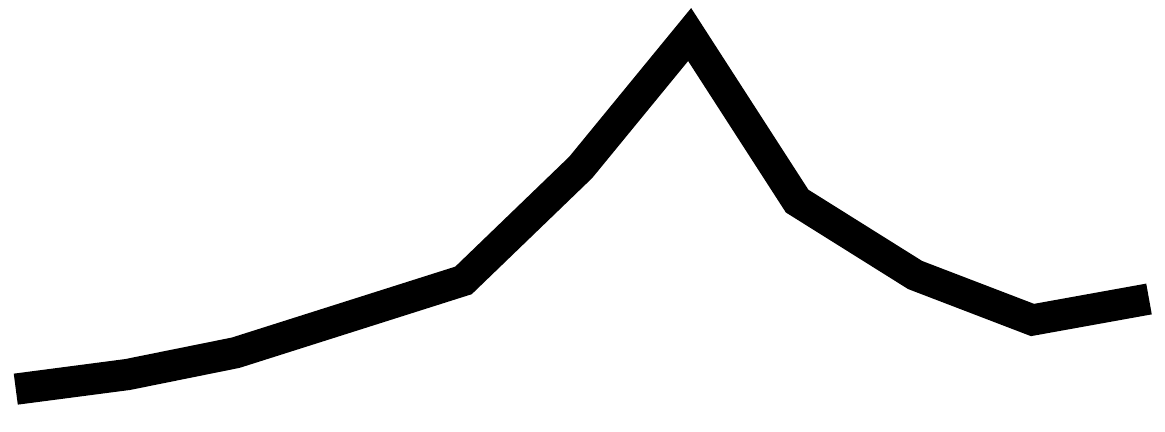}). The clusters provoked G2 to generate a hypothesis
 regarding the properties of transients:
 \textit{``Is that because all the transient groups
 get clustered together, or can I get sharp patterns
 that rise and ebb at different time points?''}
 To verify this hypothesis, G2 increased the parameter controlling the number of clusters and noticed that the clusters
 no longer exhibited the clean,
 intuitive patterns he had seen earlier.
 G3 expressed a similar sentiment and proceeded
 by inspecting the visualizations
 in the cluster via drag-and-drop.
 He found a group of genes that all transitioned
 at the same timestep, while others transitioned
 at different timesteps.\techreport{G3 described the process of using
 VQSs as doing ``detective work'' that provoked
 him to generate further scientific hypotheses
 as well as data actions.} By browsing through the ranked list of
 results\cut{(\bartext{ranked results} in Figure~\ref{fig:origins_of_sketch})}, participants were also able to gain a peripheral overview of the data and spot anomalies during exploration. For example, A1 spotted time series that were too faint to look like stars after applying the filter CLASS\_STAR=1,
 which led him to discover that all stars have been mislabeled with CLASS\_STAR=0 as 1 during data cleaning.
 \par \textbf{Context creation} in VQSs enables users to change the ``lens''
 by which they look through the data
 when performing visual querying,
 thereby creating more opportunities
 to explore the data from different perspectives. Echoing the sentiment from past studies in visual analytics regarding the importance of designing features that enable users to select relevant subsets of data~\cite{Shneiderman1994,Amar2005,Heer2012,Lee2019}, we found that all participants found at least one of the features in context creation to be useful.
 \par Both A1 and A2 expressed that context creation through interactive filtering \rchange{was a powerful way to dynamically} test conditions and tune values that they would not have otherwise \rchange{experimented with}, effectively lowering the barrier between the iterative hypothesize-then-compare cycle during sensemaking.
 During the study, participants used filtering
 to address questions such as:
 \textit{Are there more genes similar
 to a known activator when we subselect
 only the differentially expressed genes?} \techreport{\texttt{DIFFEXP=1} }(G2) and \textit{Can I find more supernovae candidates if I query only on objects that are bright and classified as a star?} \techreport{\texttt{flux\textgreater10 AND CLASS\_STAR=1} }(A1). Three participants had also used filtering as a way to query with known individual objects of interest\cut{, as shown in the \bartext{object of interest} bar of Figure~\ref{fig:origins_of_sketch}}. For example, G2 set the filter as gene=9687 and explained that since ``\textit{this gene is regulated by the estrogen receptor, when we search for other genes that resemble this gene, we can find other genes that are potentially affected by the same factors.}''
 \par While filtering enabled users to
 narrow down to a selected data subset,
 dynamic classes (buckets of data points that satisfies one or more range constraints) enabled users to compare
 relationships between multiple attributes and subgroups of data.
 For example, M2 divided solvents in the database
 into eight different categories based on voltage properties,
 state of matter, and viscosity levels,
 by dynamically setting the cutoff values
 on the quantitative variables to create these classes.
 By exploring these custom classes, M2 discovered that the relationship between viscosity and lithium solvation energy is independent of whether a solvent belongs to the class of high voltage or low voltage solvents. He cited that dynamic class creation was central to learning about this previously-unknown attribute properties:
 \begin{quote}
 All this is really possible because of dynamic class creation, so this allows you to bucket your intuition and put that together. [...] I can now bucket things as high voltage stable, liquid stable, viscous, or not viscous and start doing this classification quickly and start to explore trends. [...] look how quickly we can do it!
 \end{quote}
 \subsection{Combining Sensemaking Processes in VQS Workflows}
 Given our observations so far as to how participants make use of each sensemaking process in practice, we construct a Markov model to further investigate the interplay between these sensemaking processes in the context of an analysis workflow. \rchange{Markov models have been used in the past by Reda et al.~\cite{Reda2016} in a similar manner to analyze interaction sequences from open-ended, exploratory analysis evaluation studies. The goal of such analysis is to quantitatively capture how users ``\textit{transitions between mental, interaction, and computational states}'' to afford researchers to qualitatively characterize the processes and behavioral patterns ``\textit{essential to insight acquisition}''~\cite{Reda2016}.} 
 \par
\begin{figure}[ht!]
   \centering
   \includegraphics[width=\linewidth]{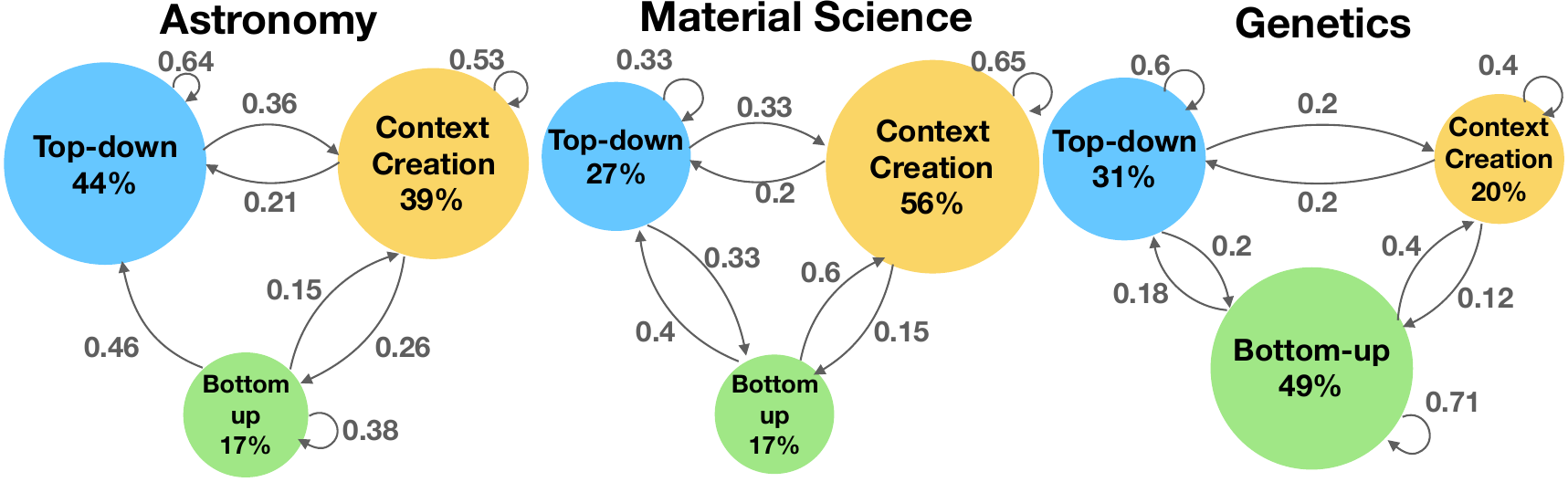}
   \caption{Markov models computed based on evaluation study event sequences, with edges denoting the probability that participant in the particular domain will go from one sensemaking process to the next. Nodes are scaled according to their eigenvector centrality, representing the percentage of time participants would spend in a particular sensemaking process in steady state. The data consists of 206 event actions taken by participants during the study (80 for astronomy, 65 for genetics, and 61 for material science).}\label{fig:transition}
   \vspace*{-20pt}
 \end{figure}

 To compute the state transition probabilities in the Markov model, we make use of event sequences from the evaluation study, where each event consists of labels describing when specific features were used. Using the taxonomy in Table~\ref{bigfeaturetable}, we map each usage of a feature in \zvpp to one of the three sensemaking processes. Each participant's event sequence is divided into sessions, each indicating a separate line of inquiry during the analysis. Based on these event sequences---one for each session, we compute the aggregate state transition probabilities (edge weight labels in Figure~\ref{fig:transition}) to characterize how participants from each domain move between different sensemaking processes\rchange{\footnote{Results were broken down by domain, rather than on an individual basis, since the analytical patterns within the domains are very similar (possibily due to the similarity between analytical inquiries and datasets within the domains).}}. 
 \par \rchange{The transition probability represents the probability that an action from one class would be followed by one from the other.} For example, in material science, \rchange{60\% of events that started with }bottom-up exploration \rchange{lead} to context creation and to top-down pattern search the rest of the time. Self-directed edges indicate the probability that the participant
 would continue with the same type of sensemaking process. For example, when an astronomer performs top-down pattern search, \rchange{64\% of the transitions were }followed by another top-down process and \rchange{by} context creation the rest of the time,
 but never followed by a bottom-up process.
 This high self-directed transition probability
 reflects how astronomers often need to iteratively
 refine their top-down query through pattern
 or match specification when looking for a specific pattern. 
 \cut{
 \begin{table}[ht!]
   \centering
   \includegraphics[width=\linewidth]{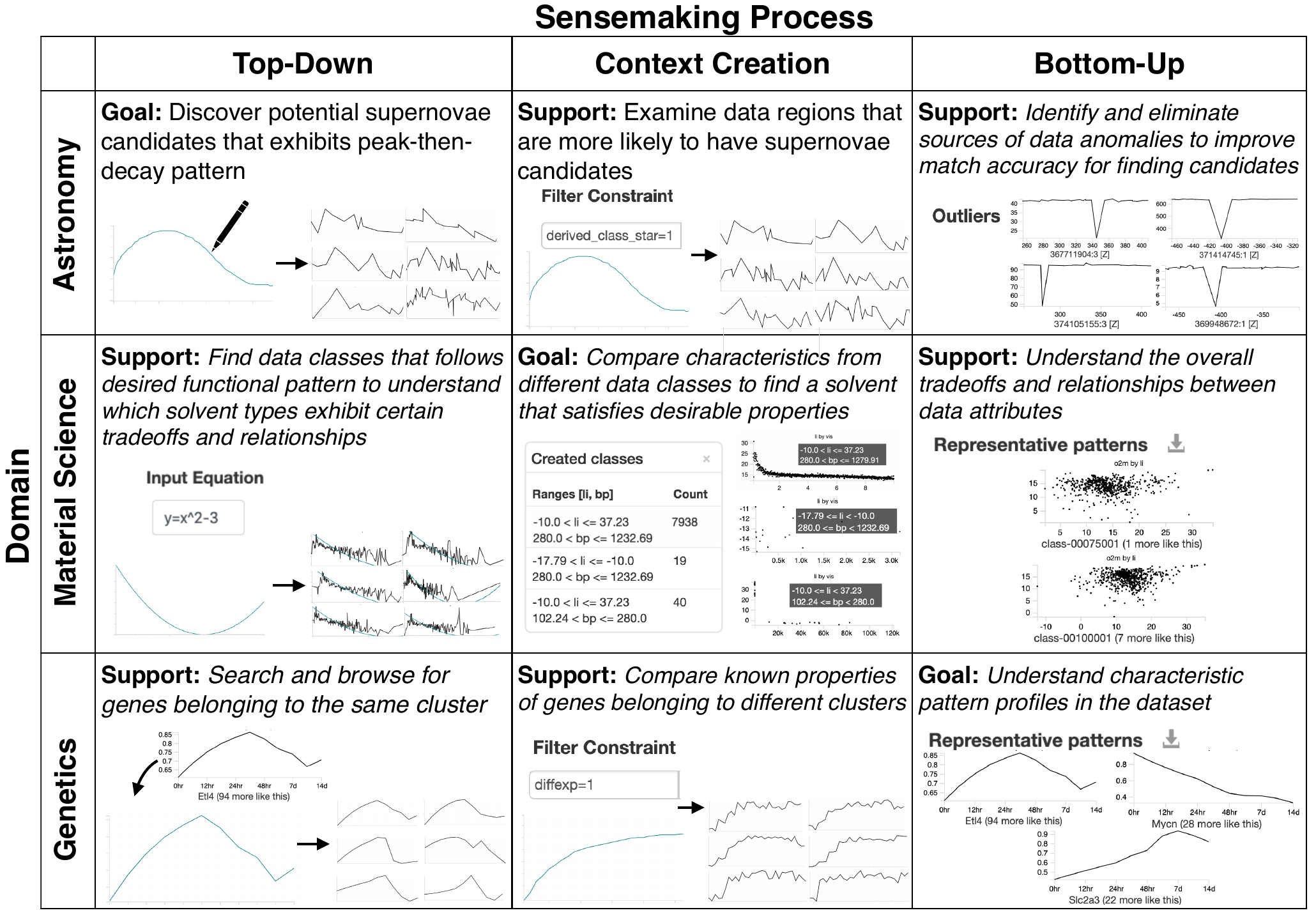}
   \caption{\rchange{Table of example usage scenarios from each domain for each sensemaking process.}\cut{Each VQS sensemaking process maps to scientific tasks and goals from each use case, from pattern search to comparing visualization collections to improving overall data understanding.} We find that our participants typically have one focused goal expressible through a single sensemaking process, but since their desired insights may not always be achievable with a single class of operation, \rchange{they make use of the two other sensemaking processes to support\cut{ (\textbf{Support})} them in accomplishing their main goal\cut{  (\textbf{Goal})}}.}
   \label{science_task}
   \vspace{-10pt}
\end{table}
}
 \par To study how important each sensemaking process
 is for participant's overall analysis,
 we compute the eigenvector centrality of each graph,
 displayed as node labels in Figure~\ref{fig:transition}.
 These values represent the percentage of time the participants
 spend in each of the sensemaking processes
 when the transition model has evolved to a steady state~\cite{pierre2011}.
 Given that nodes in Figure~\ref{fig:transition}
 are scaled by this value, in all domains,
 we observe that there is always a prominent node
 connected to two less prominent ones---but it is also clear
 that all three nodes are essential to all domains.
 Our observation demonstrates how \emph{participants
 often construct a central workflow
 around a main sensemaking process based on their analytical \textbf{goals}
 and interleave variations with the two other \textbf{support} processes as they iterate on the analytic task}. \cut{This finding is further illustrated in Table~\ref{science_task}, where the usage scenarios exemplify how each sensemaking process supports essential subtasks and enables participants' scientific goals.}
 For example, the material scientists focus on context creation 56\% of the time, mainly through dynamic class creation, followed by bottom-up inquiries (such as drag-and-drop) and top-down pattern searches (such as sketch modification).
 The central process adopted by each domain
 is tightly coupled with the problem characteristics associated with each domain. For example, without an initial query in mind,
 geneticists relied heavily on bottom-up querying
 through recommendations to jumpstart their queries.
 \par The Markov transition model exemplifies how participants
 adopted a diverse set of workflows
 based on their unique set of research questions. The bi-directional and cyclical nature
 of the transition graphs in Figure~\ref{fig:transition} highlight how the three sensemaking processes do not simply follow a linear progression towards finding a single pattern or attribute of interest. 
 Instead, the high connectivity of the transition model illustrates how these three equally-important processes form a sensemaking loop, representing iterative acts of dynamic foraging and hypothesis generation. This finding reinforces the importance of each sensemaking process and indicates that future VQSs need to be \emph{integrative} in supporting all three sensemaking process to enable a diverse set of potential workflows for addressing a wide range of analytical inquiries. 
 \subsection{Limitations}
 \par Although evidence from our evaluation study points to the infrequent use of direct sketch, we have not performed controlled studies with a sketch-only system as a baseline to validate this hypothesis. \rchange{While we employed quantitative comparisons in various analysis throughout this section, our goal is to gain a formative understanding of VQS usage behavior across our small sample. Future studies with larger sample sizes and more representative samples are required to generalize these findings.} The goal of our study is to uncover qualitative insights that might reveal why VQSs are not widely used in practice; further validation of specific findings is out of the scope of this paper. While concerns regarding study results being focused on \zvpp must be acknowledged, we note that \zvpp is one of the most comprehensive VQSs to-date, covering many of the features from past systems and more (as evident from Table~\ref{table:relatedwork}). We believe that our integrative VQS, \zvpp, can serve as a baseline for future research in VQS to evaluate against and build upon. Given that this paper covered three design studies along with one evaluation study, we were unable to cover each domain to the level of detail typically found in a dedicated design study paper. Instead, our focus was to highlight the differences and similarities among these domains relevant to the capabilities required in VQS\cut{and we defer domain-specific participatory design details and artifacts to Appendix~\ref{apdx:pdartifact}}. Future longitudinal studies may also help alleviate the novelty effects that participants may have experienced during the evaluation study. While we have generalized our findings beyond existing work by employing three different and diverse domains\cut{(see Figure~\ref{fig:transition})},
 our case studies have so far
 been focused on scientific data analysis with domain-experts,
 as a first step towards greater adoption of VQSs.
 Other potential domains that could benefit from VQSs include:
 financial data for business intelligence,
 electronic medical records for healthcare,
 and personal data for quantified self.
 These different domains may each pose different sets of challenges (such as designing for novices) unaddressed by the findings in this paper,
  pointing to a promising direction for future work.
 \section{Conclusion\label{sec:conclusion}}
 While VQSs hold tremendous promise in accelerating data exploration, they are rarely used in practice. We worked closely with analysts from three diverse domains to characterize how VQSs can address their analytic challenges, collaboratively design VQS capabilities, and evaluate how VQSs are used in practice. Participants were able to use our final system, \zvpp, for discovering desired patterns, trends, and valuable insights to address unanswered research questions. Based on these experiences, we developed a sensemaking model for how analysts make use of VQSs. Contrary to past work, we found that sketch-to-query is not as effective in practice as past work may suggest. Beyond sketching, we find that each sensemaking process fulfills a central role in participants' analysis workflows to address their high-level research objectives. We advocate that future VQSs should invest in understanding and supporting all three sensemaking processes to effectively ``close the loop'' in how analysts interact and perform sensemaking with VQSs. 
 \cut{While more work certainly remains to be done, by contributing to a better understanding of how VQSs are used in practice across domains, our paper can serve as a roadmap towards the broader adoption of VQSs for novel future use cases.}

\newpage
\bibliographystyle{abbrv-doi}
\bibliography{reference}
\clearpage
\onecolumn
 \appendix
 \npar In Appendix A, we first describe additional details about the participatory design process, as well as domain-specific artifacts collected from contextual inquiry. Next, in Appendix B, we articulate the space of problems amenable to VQSs and describe how the sensemaking processes (introduced in Section~\ref{sec:sensemaking}) fit into different parts of the problem space. In Appendix C, we provide supplementary information regarding our analysis methods and results for the evaluation study. In Appendix D, we acknowledge the individuals and agencies that have made this work possible. 
 \section{Artifacts from Participatory Design\label{apdx:pdartifact}}
\begin{figure}[h!]
 \captionsetup{font=normalsize,labelfont=normalsize}
 	\centering
 	\includegraphics[width=0.85\linewidth]{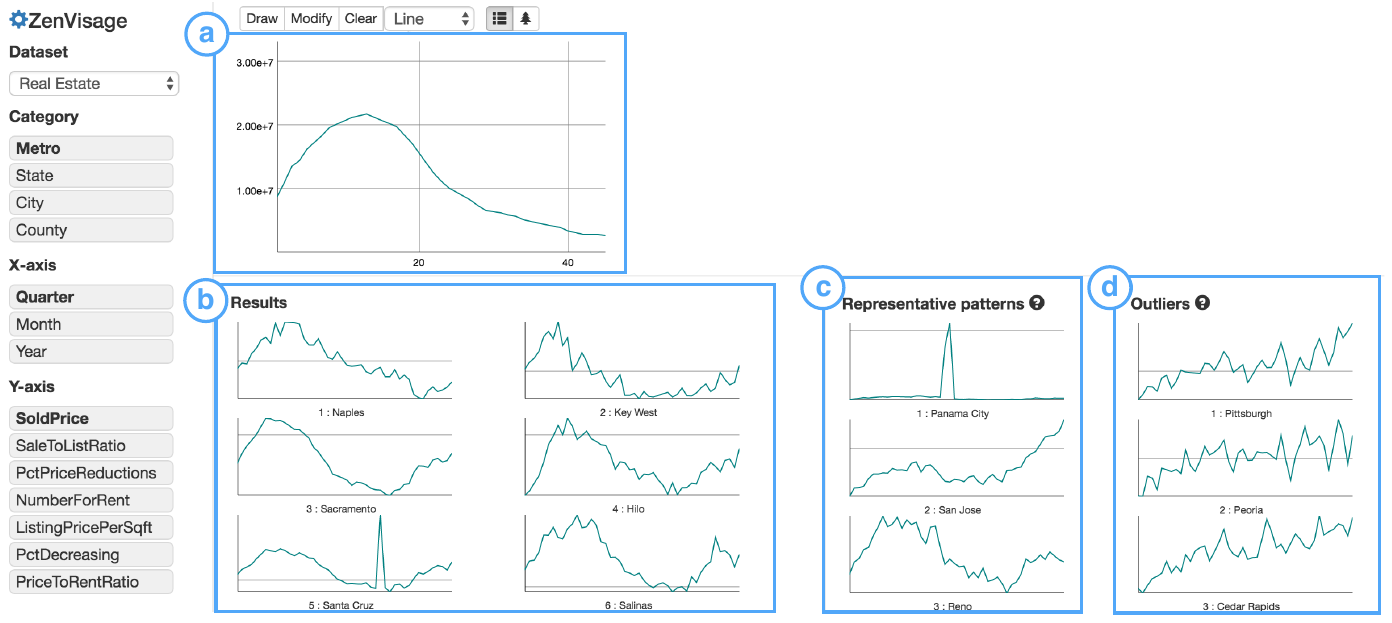}
 	\caption{The existing \zv prototype allowed users to sketch a pattern in (a), which would then return (b) results that had the closest Euclidean distance from the sketched pattern. The system also displays (c) representative patterns obtained through K-Means clustering and (d) outlier patterns to help the users gain an overview of the dataset.}
 	\label{oldZV}
 \end{figure}

\begin{figure}[h!]
    \captionsetup{font=normalsize,labelfont=normalsize}
   \centering
   \includegraphics[width=0.55\linewidth]{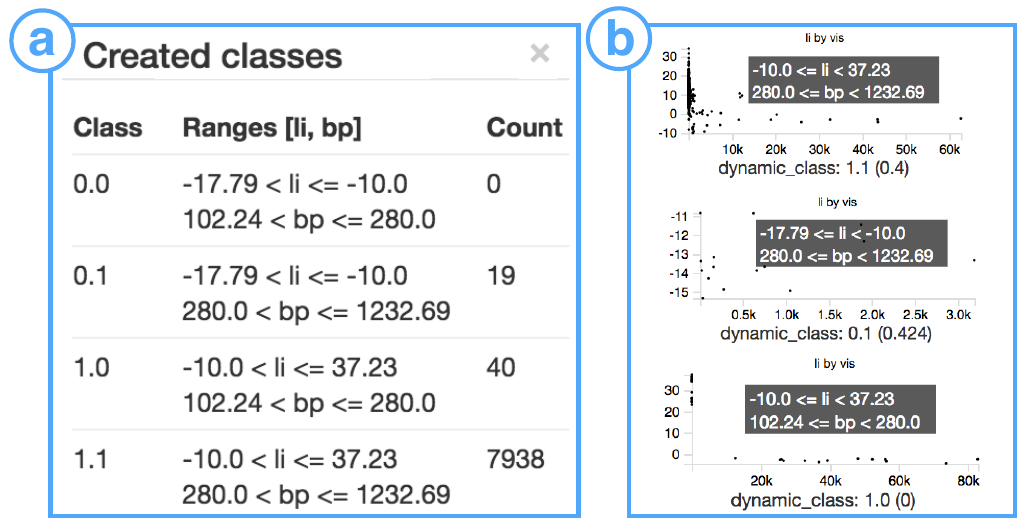}
   \caption{Example of dynamic classes. (a) Four different classes with different Lithium solvation energies (li) and boiling point (bp) attributes based on user-defined data ranges. (b) Users can hover over the visualizations for each dynamic class to see the corresponding attribute ranges for each class. The visualizations of dynamic classes are aggregate across all the visualizations that lie in that class based on the user-selected aggregation method.}
   \label{dcc}
 \end{figure}

 \npar Information about each participants can be found in Table~\ref{participants}. 
  \begin{table}[h!]
  \captionsetup{font=normalsize,labelfont=normalsize}
    \centering
    \includegraphics[width=0.5\linewidth]{figures/participant_info.pdf}
    \caption{Participant information. The Likert scale used for dataset familiarity ranges from 1 (not familiar) to 5 (extremely familiar).}
    \label{participants}
  \end{table}
 \npar During the contextual inquiry, participants demonstrated the use of domain-specific tools for conducting analysis in their existing workflow, including:
   \begin{denselist}
     \item \href{http://descut.cosmology.illinois.edu}{Image Cutout Service (Astronomy)}
     \item \href{http://cs.cmu.edu/~jernst/stem/}{Short Time-series Expression Miner (Genetics)}
     \item \href{http://srdata.nist.gov/solubility/}{Solubility Database (Material Science)}
   \end{denselist}
\begin{figure}[h!]
\captionsetup{font=normalsize,labelfont=normalsize}
 \centering
 \includegraphics[width=0.8\linewidth]{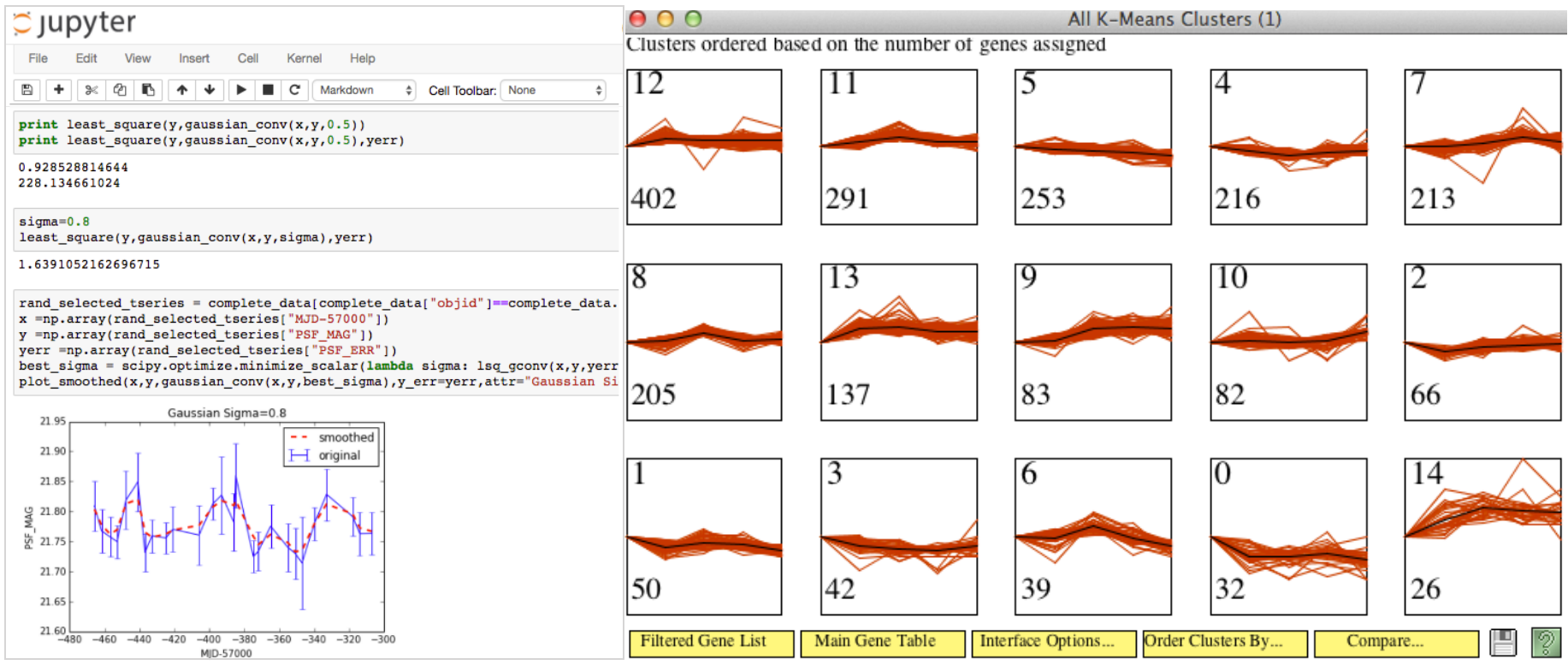}
 \caption{\rchange{Screenshots from contextual inquiry. Left: A1 performs data smoothing to clean the data and then examines a light curve manually using a Jupyter notebook. Right: G2 uses a domain-specific software to perform clustering and visualize the outputs.}}
 \label{CIscreenshot}
\end{figure}
\begin{figure}[h!]
  \captionsetup{font=normalsize,labelfont=normalsize}
    \centering
    \includegraphics[width=0.5\linewidth]{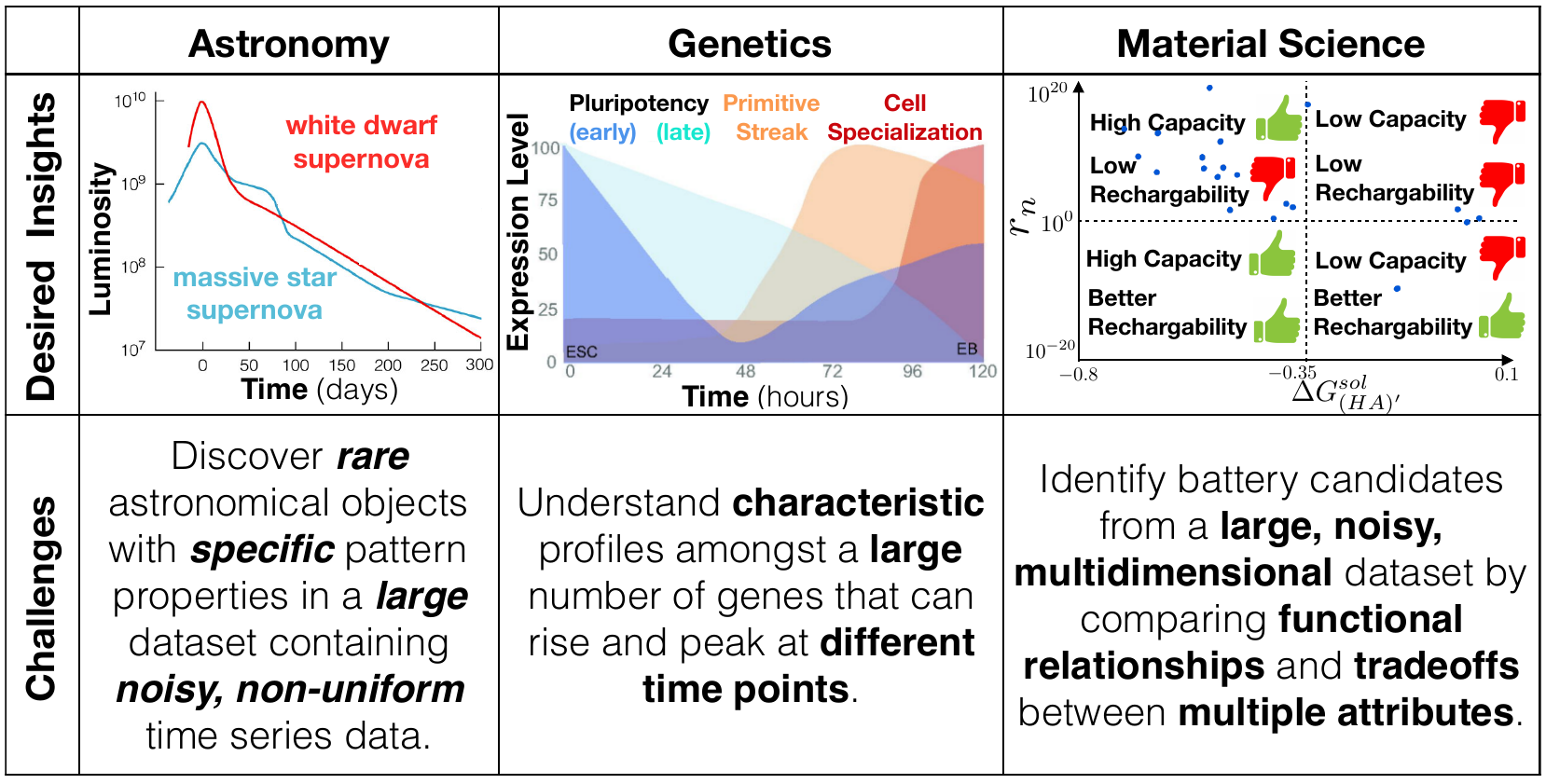}
    \caption{Desired insights, problem and dataset challenges for each of the three application domains in our study.}
    \label{science_goal}
   \end{figure}
 \newpage
 \npar Our collaboration with participants is illustrated in Figure~\ref{timeline}, where we began with an existing VQS (\zv, as illustrated in Figure~\ref{oldZV}) and incrementally incorporated features, such as dynamic class creation (Figure~\ref{dcc}), throughout the PD process. 
 \begin{figure}[h!]
 \captionsetup{font=normalsize,labelfont=normalsize}
 	\centering
   \includegraphics[width=0.8\linewidth]{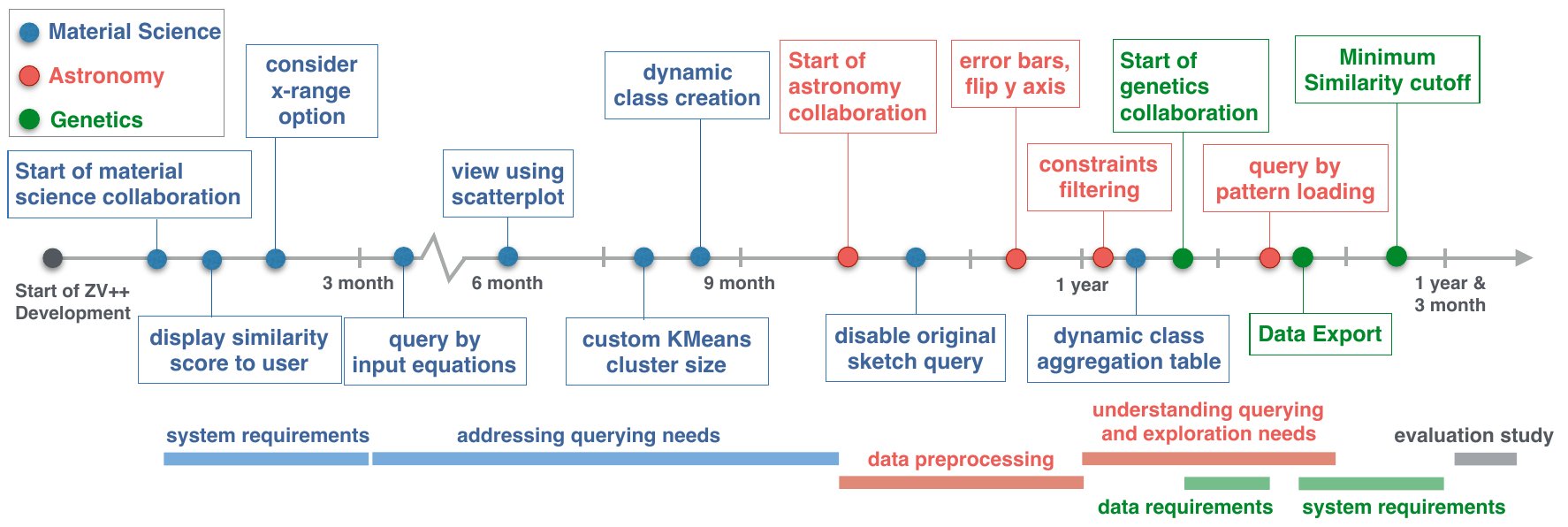}
 	\caption{Timeline for progress in participatory design studies.}
 	\label{timeline}
 \end{figure}
 \clearpage
 \section{Characterizing the Problem Space for VQSs\label{appdx:problem_space}}
\begin{figure}[h!]
 \captionsetup{font=normalsize,labelfont=normalsize}
   \centering
   \includegraphics[width=0.5\linewidth]{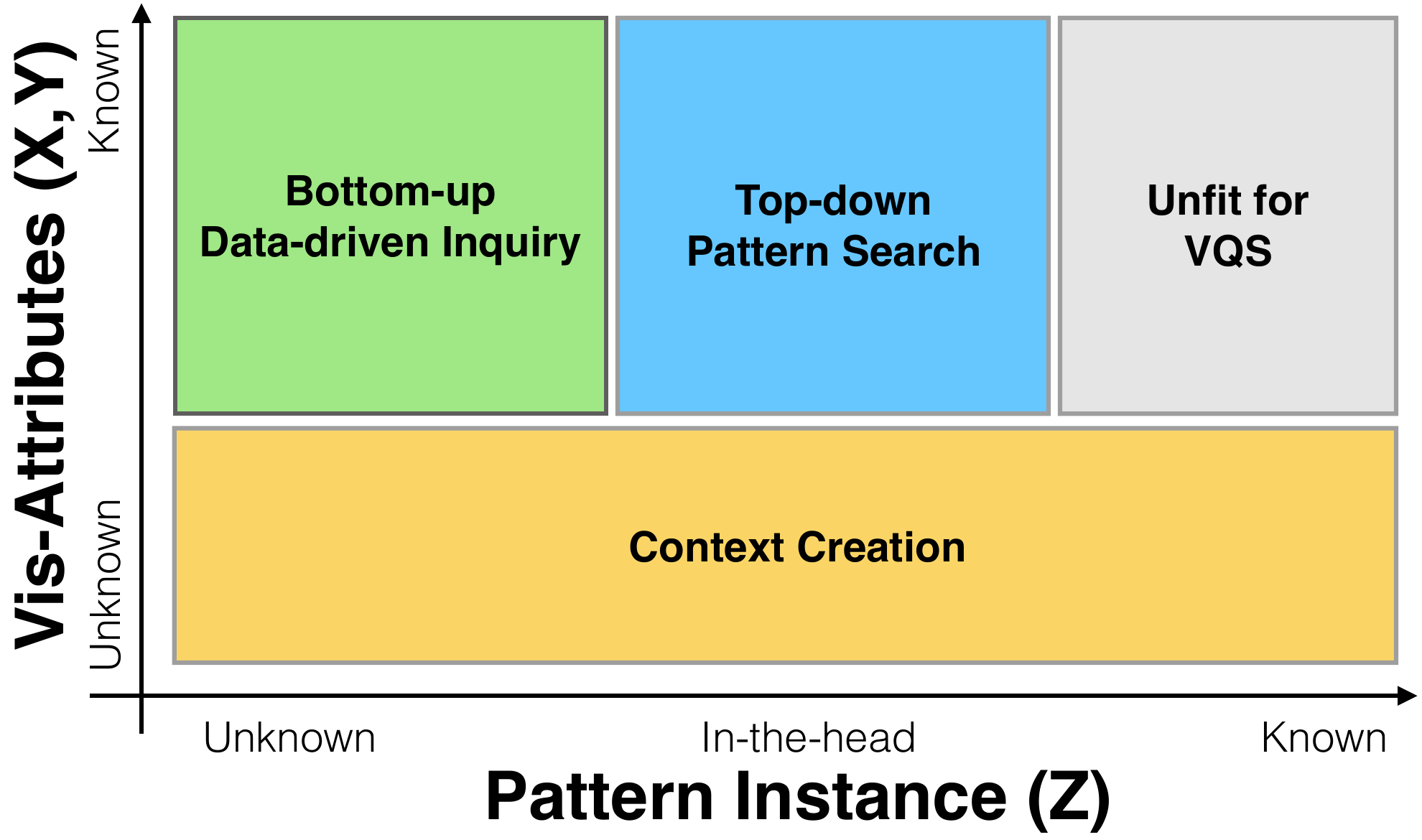}
   \caption{The problem space for VQSs is characterized by how much the analyst knows about the visualized attributes and the pattern instance. Colored areas highlight the three sensemaking processes in VQSs for addressing these characteristic problems. While prior work has focused solely on use cases in the blue region, we envision opportunities for VQSs beyond this to a larger space of use cases covered by the yellow and green regions.}
   \label{2dmodel}
 \end{figure}

 We now characterize the space of problems addressable by VQSs and describe how each sensemaking process fits into different problem areas that VQSs are aimed to solve. Visual querying often consists of searching for a desired pattern instance (Z) across a visualization collection specified by some given attributes (X,Y). Correspondingly, we introduce two axes depicting the amount of information known about the visualized attribute and pattern instance as shown in Figure~\ref{2dmodel}.
 \npar Along the \textbf{pattern instance} axis, the visualization that contains the desired pattern may already be \texttt{known} to the analyst, exist as a pattern \texttt{in-the-head} of the analyst, or be completely \texttt{unknown} to the analyst. In the \texttt{known} pattern instance region (Figure~\ref{2dmodel} grey cell), systems such as Tableau, where analysts manually create and examine each visualization one at a time, is more well-suited than VQSs, since analysts can directly work with the selected instance without having to search for which visualization exhibits the desired pattern. We define \textit{top-down pattern search} as the process where analysts query a fixed collection of visualizations based on their in-the-head pattern (Figure~\ref{2dmodel} blue). On the other hand, \textit{bottom-up data-driven inquiries} (Figure~\ref{2dmodel} green) are driven by recommendations or queries that originate from the data (or equivalently, the visualization), since the pattern of interest is unknown and external to the user.
 \npar The second axis, \textbf{visualized attributes},
 depicts how much the analyst
 knows about which X and Y axes
 they are interested in visualizing.
 In both the astronomy and genetics use cases,
 as well as past work in this space, the attribute to be visualized is \texttt{known}, as data was in the form of a time series. In the case of our material science participants, they wanted to explore relationships between different
 X and Y variables. In this realm of \texttt{unknown} attributes, context creation (Figure~\ref{2dmodel} yellow) is
 essential for allowing users to pivot across different visualization collections.
 \rchange{
  \clearpage
   \section{Evaluation Study Protocol\label{apdx:studyprotocol}}
   Here, we detail the procedures that were conducted during the evaluation study. At the beginning of the study, participants were asked a set of pre-study survey questions to collect basic information about participant's dataset, scientific questions, and existing workflows. While this information was similar to the ones collected through participatory design and contextual inquiry (Section~\ref{sec:participantdatasets}), the pre-study survey ensured that we have background information even for the ``blank-slate'' participants (who were not part of the earlier design study).
   \begin{itemize}
    \item What is your current role as a scientist? What are some examples of recent questions you have researched?
    \item Describe the workflow that you currently use to analyze and make sense of this type of data.
    \item Can you describe an interesting finding you found with your current workflow and the process you took to obtain this insight? 
   \end{itemize}
   \smallskip\par\noindent After the tutorial and overview of the system, participant's selected dataset was loaded in. Participants were asked about their familiarity with the dataset and their analytical goals for the session.
   \begin{itemize}
    \item On a scale of 1-5, how familiar are you with this dataset? How long have you been working with this dataset? If you have worked with this dataset before, is there any insight that you already know from this dataset? 
    \item What is your goal for this dataset? What are you hoping to accomplish with this dataset?
   \end{itemize}
   \smallskip\par\noindent During the main experiment, participants engaged in talk-aloud exercises as they explored their data. These two semi-structured interview questions were often posed when participants begin a new line of analytical inquiry.
   \begin{itemize}
    \item What is your current goal in this phase of the exploration? What type of insights are you hoping to obtain? 
    \item What actions are you planning to perform? How are you operationalize to achieve those goals?
   \end{itemize}
   In addition, we occasionally remind participants that they ask for help on something they want to accomplish on \zvpp, but were not sure about the sequence of interactions. They were also encouraged to use other tools in their existing workflow alongside \zvpp to perform their analysis. 
   \smallskip\par\noindent At the end of the study, we interviewed participants with a set of open-ended questions regarding their experience with \zvpp, including:
   \begin{itemize}
     \item How was \zvpp different from your existing workflow? 
     \item Can you describe how you would use \zvpp in your current workflow?
     \item On a scale of 1-10, how interested would you be in adopting this tool for your day-to-day workflow?
     \item What were some insights that you have gained from today's session?
     \item Given the insights that you have obtained from \zvpp, are there any additional analysis that you will run downstream before you publish these results? Describe these additional downstream analysis steps.
     \item What are the pros and cons for using \zvpp?
     \item Were there any queries that you were unable to address with \zvpp during today's session?
     \item What are additional features in \zvpp that would help with your scientific workflow or serve your scientific need?
   \end{itemize}
}
\clearpage
 \section{Evaluation Study Analysis Details\label{apdx:studydetails}}
\begin{figure}[h!]
 \captionsetup{font=normalsize,labelfont=normalsize}
     \centering
     \includegraphics[width=0.45\columnwidth]{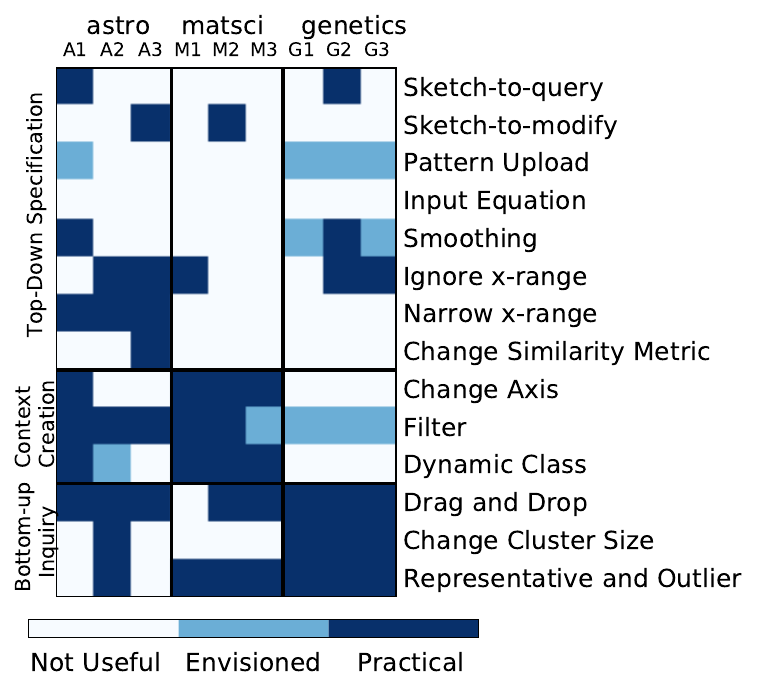}
     \vspace{-6pt}\caption{Heatmap of features categorized as practical usage (P), envisioned usage (E), and not useful (N). Columns are arranged in the order of subject areas and the features are arranged in the order of the three foraging acts. Participants preferred to query using bottom-up methods such as drag-and-drop over top-down approaches such as sketching or input equations. Participants found that context creation via filter constraints and dynamic class creation were powerful ways to compare between subgroups or filtered subsets.}
     \label{feature_heatmap}
 \end{figure}

 We analyzed the transcriptions of the evaluation study recordings through open-coding and
 categorized every event in the user study using the following coding labels:
 \begin{denselist}
     \item Insight (Science) \textbf{[IS]}: Insight that connected back to the science (e.g. ``This cluster resembles a repressed gene.'')
     \item Insight (Data) \textbf{[ID]}: Data-related insights (e.g. ``A bug in my data cleaning code generated this peak artifact.'')
     \item Provoke (Science) \textbf{[PS]}: Interactions or observations that provoked a scientific hypothesis to be generated.
     \item Provoke (Data) \textbf{[PD]}: Interactions or observations that provoked further data actions to continue the investigation.
     \item Confusion \textbf{[C]}: Participants were confused during this part of the analysis.
     \item Want \textbf{[W]}: Additional features that participant wants, which is not currently available on the system.
     \item External Tool \textbf{[E]}: The use of external tools outside of \zvpp to complement the analysis process.
     \item Feature Usage \textbf{[F]}: One of the features in \zvpp was used.
     \item Session Break \textbf{[BR]}: Transition to a new line of inquiry.
 \end{denselist}
 \begin{table}[h!]
 \captionsetup{font=normalsize,labelfont=normalsize}
  \centering
   \begin{tabular}{lrrrrrrrrr}
   \hline
    Domain           &   IS &   ID &   PS &   PD &   C &   W &   E &   BR &   F \\
   \hline
    astro            &    4 &   12 &   13 &   57 &   2 &  18 &  20 &   22 &  67 \\
    genetics         &    8 &   12 &    7 &   35 &   4 &  13 &   1 &   21 &  52 \\
    mat sci          &   14 &    8 &    7 &   44 &   8 &  11 &   3 &   12 &  48 \\
   \hline
   \end{tabular}
   \caption{Count summary of thematic event code across all participants of the same domain.}
 \end{table}
 \npar In addition, based on the usage of each feature during the user study, we categorized the features into one of the three usage types:
 \begin{denselist}
     \item Practical \textbf{[P]}: Features used in a sensible and meaningful way.
     \item Envisioned usage \textbf{[E]}: Features which could be used practically if the envisioned data was available or if they conducted downstream analysis, but was not performed due to the limited time during the user study.
     \item Not useful \textbf{[N]}: Features that are not useful or do not make sense for the participant's research question and dataset.
 \end{denselist}
 The feature usage labels for each user is summarized in Figure~\ref{feature_heatmap}. A feature is regarded as \emph{useful} if it has a \textbf{P} or \textbf{E} code label. Using the matrix from Figure~\ref{feature_heatmap}, we compute the percentage of useful features for each sensemaking process as: $\frac{\textrm{\# of useful features in process}}{\textrm{total \# of features in process} \times \textrm{total \# of users}}$.
 \vspace{-5pt}
 \begin{figure}[h!]
  \captionsetup{font=normalsize,labelfont=normalsize}
    \centering
   \includegraphics[width=0.55\linewidth]{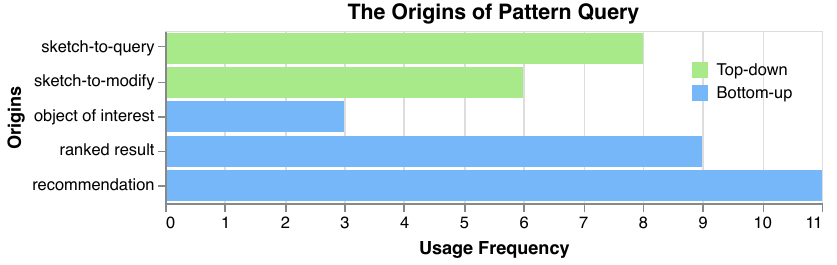}
   \vspace{-5pt}
   \caption{The number of times a pattern query originates from one of the workflows. Pattern queries are far more commonly generated via bottom-up than top-down processes.}\label{fig:origins_of_sketch}
 \end{figure}
 \begin{table}[h!]
  \captionsetup{font=normalsize,labelfont=normalsize}
   \centering
   \includegraphics[width=0.85\linewidth]{figures/science_task_new.pdf}
   \caption{\rchange{Table of example usage scenarios from each domain for each sensemaking process.}\cut{Each VQS sensemaking process maps to scientific tasks and goals from each use case, from pattern search to comparing visualization collections to improving overall data understanding.} We find that our participants typically have one focused goal expressible through a single sensemaking process, but since their desired insights may not always be achievable with a single class of operation, \rchange{they make use of the two other sensemaking processes to support\cut{ (\textbf{Support})} them in accomplishing their main goal\cut{  (\textbf{Goal})}}.}
   \label{science_task}
  \cchange{
  \section{Acknowledgments}
  \begin{flushleft}
  \npar We thank Chaoran Wang, Edward Xue, and Zhiwei Zhang, who have contributed to the development of \zvpp, as well as our scientific collaborators, who provided valuable feedback during the design study. We appreciate the constructive feedback from the anonymous reviewers, which significantly improved the quality of this paper. We acknowledge support from grants IIS-1513407,IIS-1633755, IIS-1652750, and IIS-1733878 awarded by the National Science Foundation, and funds from Microsoft, 3M, Adobe, Toyota Research Institute, Google, and the Siebel Energy Institute. The content is solely the responsibility of the authors and does not necessarily represent the official views of the funding agencies and organizations.
  \end{flushleft}
  }
\end{table}

\end{document}